\documentclass[twocolumn, prx, superscriptaddress, floatfix, nofootinbib,10pt]{revtex4-2}

\pdfoutput=1

\usepackage[utf8]{inputenc}
\usepackage[english]{babel}
\usepackage[sort&compress]{natbib}

\usepackage{graphicx}
\usepackage{amsmath, amssymb, amsthm}
\usepackage{dsfont}
\usepackage[dvipsnames]{xcolor}
\usepackage{hyperref}
\hypersetup{
   colorlinks = true,
   allcolors = RoyalPurple
}

\usepackage{quantikz}

\usepackage[export]{adjustbox}
\usepackage{BOONDOX-cal} 
\usepackage{orcidlink}
\usepackage[normalem]{ulem}
\usepackage{soul}

\newcommand{\abs}[1]{\left\lvert #1 \right\rvert}

\renewcommand{\epsilon}{\varepsilon}

\renewcommand{\braket}[1]{\langle #1\rangle}
\renewcommand{\ket}[1]{\vert #1 \rangle}
\renewcommand{\bra}[1]{\langle #1 \vert}

\begin{document}

\title{Thermodynamic-limit dispersion relations on trapped-ion quantum hardware}

\author{Lucas Marti \orcidlink{0009-0005-3130-380X}}
\email[]{lucas.marti@fau.de}
\affiliation{Department of Physics, Friedrich-Alexander Universität Erlangen-Nürnberg (FAU), Staudtstraße 7, 91058 Erlangen}

\author{Sumeet \orcidlink{0000-0002-1501-542X}}
\email[]{sumeet.sumeet@fau.de}
\affiliation{Department of Physics, Friedrich-Alexander Universität Erlangen-Nürnberg (FAU), Staudtstraße 7, 91058 Erlangen}
\date{\today}
\author{Stefan Wolf \orcidlink{0009-0003-8297-3745}}
\author{K. P. Schmidt \orcidlink{0000-0002-8278-8238}}
\affiliation{Department of Physics, Friedrich-Alexander Universität Erlangen-Nürnberg (FAU), Staudtstraße 7, 91058 Erlangen}
\author{Michael J. Hartmann \orcidlink{0000-0002-8207-3806}}

\affiliation{Department of Physics, Friedrich-Alexander Universität Erlangen-Nürnberg (FAU), Staudtstraße 7, 91058 Erlangen}
\affiliation{Quint Computing GmbH, 91058 Erlangen, Germany}
\date{\today}

\begin{abstract}
We run a numerical linked-cluster expansion with a quantum algorithm (NLCE+QA), computing ground-state energies and one quasi-particle dispersions in the thermodynamic limit using a 20-qubit trapped-ion quantum processing unit (QPU). The NLCE+QA framework extracts thermodynamic-limit properties from small-cluster calculations, making it naturally suited for near-term quantum devices. Projector-based block-diagonalization schemes such as projective cluster-additive transformation (PCAT) are essential to NLCE+QA, and they involve matrix inversion and square root operations that amplify measurement noise. A central question is therefore whether current hardware can provide expectation values that are accurate enough to withstand non-linear classical post-processing. We explore this challenge for the transverse-field Ising model (TFIM) in one dimension, on a ladder geometry, as well as in a longitudinal field in one dimension. For the quantum algorithm, we consider adiabatic state preparation (ASP), as well as a variational quantum eigensolver (VQE) trained on a classical device. The final expectation values are obtained from the QPU, using a novel alternative to the Hadamard test that we name the CX-test. We explore the regimes currently attainable on quantum devices and comment on the improvements needed for quantum computers to achieve results beyond classical reach.
\end{abstract}

\maketitle

\section{Introduction}

Computing properties of quantum many-body systems in the thermodynamic limit is of central interest in condensed matter physics, yet classical methods face fundamental limitations for strongly correlated systems. Quantum computers offer a promising way forward, but current devices remain constrained in both qubit number and gate fidelity, restricting direct calculations to small system sizes far from the thermodynamic limit. 

Methods like numerical linked-cluster expansions (NLCEs) \cite{nlce_Rigol, TANG2013557} bridge this gap by providing a systematic route to estimating properties of large systems in the thermodynamic limit from calculations on small clusters alone. The method decomposes extensive quantities into contributions from connected clusters via an inclusion-exclusion principle, and is naturally suited for integration with current quantum hardware, since only small clusters serve as input. While dynamical mean-field theory (DMFT) also approximates the thermodynamic limit, single-site DMFT lacks spatial correlations and is best suited to high-dimensional lattices~\cite{cluster-dmft,wolf2025}; cluster extensions partially restore short-range correlations but remain limited in scope. Tensor network methods such as infinite projected entangled pair states (iPEPS)~\cite{ipeps} provide an alternative route to thermodynamic-limit properties that naturally incorporate spatial correlations.

Our approach combines NLCEs with quantum algorithms, where either ASP~\cite{albash_Adiabatic_2018} or a VQE~\cite{Peruzzo2014} can be used as the quantum algorithm, in the second case resulting in a hybrid quantum-classical framework. Ref.~\cite{sumeet2024} introduced the NLCE+VQE method, in which a classical simulation of this method was applied to calculate the ground-state energy of the TFIM in the thermodynamic limit. In a subsequent work~\cite{sumeet2025}, this approach was extended to compute quasi-particle excitation energies. Unlike ground-state energies, excitation energies are encoded in a matrix rather than a single scalar quantity. This introduces additional hurdles. In particular, ensuring cluster additivity becomes a central challenge, which we ensure by the projective cluster-additive transformation (PCAT)~\cite{Hormann2023a}.

In this work, we present the first hardware implementation of the NLCE+VQE framework proposed in~\cite{sumeet2024,sumeet2025}, and introduce the NLCE+ASP framework, computing one quasi-particle (1QP) dispersions in the thermodynamic limit on a 20-qubit trapped-ion QPU. A central challenge in bringing this method to real hardware revolves around the capacity of current devices to provide expectation values that are accurate enough for the non-linear classical post-processing inherent in projector-based block-diagonalization schemes such as PCAT, which involves matrix inversion and matrix square root operations. This non-linear structure is not specific to PCAT but is shared by other block-diagonalization approaches such as the Schrieffer-Wolff (SW) transformation; PCAT additionally guarantees cluster additivity, which is required for the NLCE to converge. More broadly, non-linear functions of multiple expectation values pose a generic challenge for noisy quantum devices. On classical computers, operations such as matrix inversions~\cite{friedrich_Precision_2018} or fractional powers~\cite{lischke_What_2019} are routine and arise naturally across many areas of physics~\cite{lakes_Viscoelastic_2009,romano_Fundamental_2025}, and thus must be part of any simulation thereof. On noisy quantum hardware, however, the sensitivity of non-linear functions to noise~\cite{lefebvre_Propagation_2000} makes their use far less straightforward. Similar questions arise in classical computations that have a stochastic component, for instance variational Monte Carlo~\cite{Carleo17,Hartmann19}.

Non-linear processing of quantum data has already been explored in other contexts. In a recent work by \textcite{schiffer_Hardwareefficient_2025}, a Hadamard test is performed iteratively, and the final value is obtained by classically computing a product of fractions 
$g = \frac{g_{n}}{g_{n-1}} \cdots \frac{g_{1}}{g_{0}}$, where all $g_j$, $j \in 1,\ldots,n$, are expectation values obtained from the quantum computer. In such a non-linear combination, the propagation of uncertainty must be carefully controlled, especially when the $g_j$ are small. Nonetheless, this example finely illustrates that combining expectation values non-linearly can bring out novel algorithmic strategies. It is then natural to ask whether current QPUs have low enough error rates to be integrated into algorithms that make systematic use of non-linear post-processing, while also being able to sample enough to collect sufficient statistics.

We study the challenges that arise when computing the 1QP dispersion in the thermodynamic limit on a QPU, using the NLCE+QA procedure, and mitigate them by introducing a novel alternative to the Hadamard test. Our results, obtained for the polarized phase of the TFIM in one dimension, in a longitudinal field, and on the ladder geometry, indicate that, in limited regimes, the non-linear post-processing pipeline is viable on current hardware, producing recognizable dispersion relations that approach the expected behavior. 

Our work opens the door to thermodynamic-limit computation with spatial correlations performed on a QPU, and it addresses a relevant problem in condensed matter physics, for which quantum devices could yield advantages over classical methods. It also indicates that quantum hardware precisions have reached a stage where it becomes meaningful to seek new algorithms that involve non-linear classical post-processing of quantum data.

We note that there is an important distinction between non-linear classical post-processing of sets of expectations, and non-linear functions of quantum states. The former is a hybrid quantum-classical scheme, suited for noisy devices, while the latter directly reflects a physical quantity. The simplest example of these is von Neumann entropy, which can be computed, for example, with block-encodings~\cite{wang_New_2024}, or classical shadows~\cite{huang_Predicting_2020}. 

As hardware rapidly improves and the transition between NISQ and fault-tolerance~\cite{eisert_Mind_2025} broadens the scope of possible algorithms, hybrid approaches pairing quantum measurements with non-linear classical post-processing offer a promising path forward.

The remainder of this work is organized as follows. In Sec.~\ref{sec:dispersion}, we describe the theoretical framework for computing the 1QP dispersion on quantum hardware, including the block-diagonalization strategies (Sec.~\ref{sec:framework}), the PCAT construction and non-linear post-processing (Sec.~\ref{subsec:PCAT}), the measurement schemes for matrix elements (Sec.~\ref{sec:measurements}), and the NLCE embedding (Sec.~\ref{subsec:nlce}). In Sec.~\ref{sec:error_analysis}, we discuss the effect of shot and hardware noise on the results. In Sec.~\ref{sec:results}, we present the hardware results for the different TFIM models that we picked. We finally discuss future work in Sec.~\ref{sec:outlook}.

\section{Approach}\label{sec:dispersion}

Our goal is to compute the 1QP dispersion $\omega(\vec{k})$ of a quantum spin Hamiltonian of the form $H = H_0 + xV$, in the thermodynamic limit on quantum hardware using a numerical linked-cluster expansion. Here, $\vec{k}$ is the momentum of the quasi-particle and the $H_0$ term is exactly solvable with a non-degenerate ground state. It can thus be written in a block-diagonal form with respect to QP sectors. To find $\omega(\vec{k})$, we implement a unitary $U$ that block-diagonalizes $H$ for $x>0$, decoupling the 1QP sector from all other quasi-particle sectors. When doing so, we assume that the dressed QP at finite $x$ is adiabatically connected to the undressed limit $x\rightarrow 0$, i.e., there is no quantum phase transition along the way and no quasi-particle decay occurs at any momentum. 

The NLCE framework allows for the determination of physical quantities in the thermodynamic limit by performing calculations on finite connected clusters which are appropriately embedded in the infinite system. In the following we first present all relevant information for the treatment of clusters using quantum algorithms and then turn to explain the NLCE scheme.

\subsection{Decoupling 1QP using quantum algorithms on clusters}
\label{sec:framework}

In this subsection, we discuss how to implement quantum algorithms as cluster solvers within the NLCE framework on quantum hardware, including the block-diagonalization strategies, the non-linear post-processing required by the PCAT, and the measurement schemes needed to obtain the necessary matrix elements from the quantum device.

In what follows, all quantities are therefore defined on a single finite cluster $C$ with $N$ spins. The extension to the thermodynamic limit via the NLCE is described in Sec.~\ref{subsec:nlce}. 

We consider the Hamiltonian of the form $H = H_0 + xV$. Let us stress again that the perturbation $xV$ is assumed to be small enough so that no phase transition and no quasi-particle decay has occurred for smaller values of the perturbation in the thermodynamic limit, i.e.~the 1QP excitations always correspond to dressed spin-flip excitations above the paramagnetic ground state. On a finite cluster $C$, we exclude the situation of level crossings of the 0QP and 1QP states with states having a larger QP number. 

A block-diagonalizing transformation $U$ satisfies
\begin{equation}
    H_{\text{eff}} = U^\dagger H U = \bigoplus_n H_{\text{eff}}^{[nn]},
    \label{eq:block_diag}
\end{equation}
where each block $H_{\text{eff}}^{[nn]}$ acts within the $n$QP sector. The effective 1QP Hamiltonian is then
\begin{equation}
    H_{\text{eff}}^{[1]} = H_{\text{eff}}^{[11]} - E^{[0]},
    \label{eq:1qp_hamiltonian}
\end{equation}
where $E^{[0]}$ is the ground-state energy. Subtracting $E^{[0]}$ removes the extensive ground-state contribution:
for disconnected clusters $A \cup B$, a 1QP excitation on $B$ carries total energy $E^{[0]}_A + E^{[1]}_B$ and subtracting $E^{[0]}= E_A^{[0]} + E_B^{[0]}$ yields the excitation energy on B alone. This makes $H^{[1]}_{\text{eff}}$ an intensive quantity, encoding the one quasi-particle excitation energy which does not scale with system size. Without this subtraction, the reduced contribution of a disconnected cluster $A \cup B$ would not vanish, and the NLCE would diverge when taking increasingly larger clusters into account.

\subsubsection{Quantum algorithms}

One way to realize such a block-diagonalizing unitary is via a (quasi) adiabatic transform that mixes states within the same quasi-particle sector but not across different sectors. For an undressed Hamiltonian $H_0= \sum_j Z_j$, the 1QP sector corresponds to spin-flip excitations, i.e.~computational basis states having Hamming weight one, and is therefore trivially block-diagonal. In the full Hamiltonian $H$, the 1QP sector no longer consists of simple computational basis states with Hamming weight one. Instead, each dressed 1QP state is a superposition that includes contributions from states with higher Hamming weights, reflecting the quantum fluctuations introduced by the perturbation $xV$. However, under our assumptions, it can be identified entirely through the adiabatic connection and we can define the dressed 1QP sector to be the set of states that are adiabatically connected to the 1QP states of the undressed Hamiltonian $H_0$, which we already know.

As we have just described, the block-diagonalization can be most directly implemented on a quantum computer via ASP. Given two adiabatically-connected eigenstates, $\ket{\Phi_i^{[1]}}$ of $H_0$ and  $\ket{\Psi^{[1]}_i}$ of $H$, ASP performs a unitary transform between the two,
\begin{align}
 \ket{\Psi^{[1]}_i} = \lim_{T \rightarrow \infty}  U_{\rm ASP}(T)\ket{\Phi_i^{[1]}}.
\end{align}
Its accuracy depends on the runtime $T$ of the overall algorithm, which in turn is governed by the minimum energy gap to the nearest non-degenerate eigenstate along the path~\cite{albash_Adiabatic_2018}. If the time lower bound is not respected, the sweep will leak into nearby states. In our case, the PCAT does not require the final state to be a perfect eigenstate, but a combination of the eigenstates of the correct sector, so that we may rather use
 \begin{equation}\label{eq:superposed_1qp}
    \ket{\chi_i} = U_{\rm ASP}(\tilde{T}) \ket{\Phi_i^{[1]}} = \sum_{i\ell}     \alpha_{i\ell} \ket{\Psi^{[1]}_{\ell}},
 \end{equation}
where $\tilde{T}$ is a shorter time that allows mixing 1QP states but not sectors; this somewhat relaxes the runtime constraint.

On a digital device, the sweep unitary $U_{\rm ASP}$ is discretized in time steps, where each time step is typically decomposed in Trotter steps~\cite{suzuki_Decomposition_1985}, and the Trotter step unitaries can be implemented via a standard Pauli exponential decomposition~\cite{nielsen_Quantum_2010}. 

Current quantum computers do not have error rates that are low enough to sustain algorithms as deep as ASP whose runtime respects the inverse gap condition~\cite{albash_Adiabatic_2018}. QPU results of the sweep, shown in Fig.~\ref{fig:TFIM_30j_0hl_results_sweep}, hint that the NLCE+ASP pipeline is not yet attainable, and we must find shallower circuits. We describe the case of ASP in detail in App.~\ref{app:goodsweep}, and furthermore estimate the noise regime in which ASP becomes viable.

An alternative approach to NLCE+ASP is NLCE+VQE \cite{sumeet2024,sumeet2025}. In this approach, we want to find a variational circuit $U_{\text{VQE}}(\theta)$ that approximately block-diagonalizes $H$ by decoupling the ground-state and 1QP sector from higher excitations.

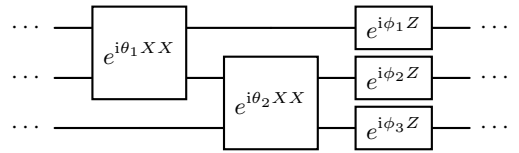
\begin{figure}[h!]
    \centering
        \begin{quantikz}[row sep=0.1cm]
        \lstick{$\cdots$} & \gate[wires=2]{e^{\mathrm{i}\theta_1 XX}} & & \gate{e^{\mathrm{i}\phi_1 Z}} & \rstick{$\cdots$} \\ 
        \lstick{$\cdots$} &  & \gate[wires=2]{e^{\mathrm{i}\theta_2 XX}}  & \gate{e^{\mathrm{i}\phi_2 Z}}&\rstick{$\cdots$} \\  
        \lstick{$\cdots$} & & & \gate{e^{\mathrm{i}\phi_3 Z}} & \rstick{$\cdots$}\\  
        \end{quantikz}
    \caption{A layer of the HVA Ansatz, for a 3-qubit TFIM chain without the longitudinal field. }
    \label{fig:hvaansatz}
\end{figure}

The variational circuit we use is based on the Hamiltonian variational ansatz (HVA)~\cite{wecker_progress_2015}, which incorporates the structure of the problem Hamiltonian. For a Hamiltonian decomposed as $H = \sum_j \alpha_j P_j$, where $P_j$ are Pauli strings and $\alpha_j$ the corresponding coefficients, the HVA applies each term as a parametrized rotation, leading to the unitary,
\begin{equation} \label{eq:hva}
    U_\mathrm{HVA}(\theta)
    = \prod_{l=1}^{L} \prod_j
      \exp\left({\rm i}\, \theta_{j,l}\, P_j\right).
   \end{equation}
A single layer consists of a sequential application of the exponentials $\exp({\rm i}\,\theta_{j,l}\,P_j)$ for each term $P_j$ in the Hamiltonian (see for instance Fig.~\ref{fig:hvaansatz}) and the full ansatz is obtained by repeating $L$ layers with independent parameters. This construction is motivated by the Trotter decomposition of an adiabatic sweep: each layer resembles a discrete time step of the sweep, with the variational parameters replacing the fixed schedule. Unless otherwise noted, all the VQEs in this work make use of the HVA ansatz with $\lceil N/2 \rceil$ layers, which was found to be the optimal number for the TFIM on one-dimensional chain and square lattice~\cite{sumeet2024}. The unitary $U_{\rm VQE}$ is then given by $U_\mathrm{HVA}$ with the optimized parameters, $U_{\rm VQE} = U_\mathrm{HVA}(\theta_{\rm opt})$. 

The NLCE+VQE implementation for 1QP requires a cost function that drives $U_\mathrm{HVA}$ towards block-diagonalization. We use the energy variance over the 1QP sector. The key property is that for any quantum state $|\psi\rangle$, the energy variance $\langle H^2 \rangle - \langle H \rangle^2$ vanishes if and only if $|\psi\rangle$ is an eigenstate of $H$, or more generally, lies within a degenerate eigenspace \cite{sumeet2025}. For our purpose, we do not need $U_{\rm HVA}$ to fully diagonalize $H$ within the 1QP sector, but only to decouple the 1QP sector from the rest of the spectrum. To this end, we minimize the total variance summed over all transformed 1QP basis states through a cost function defined as,
\begin{align} \label{eq:cf_variance}
C_{\mathrm{var}}^{1\mathrm{QP}}(\theta)
  &= \sum_{i=1}^{N} \langle \Phi^{[1]}_{i}|\,U_{\rm HVA}^\dagger(\theta)\,H^{2}\,U_{\rm HVA}(\theta)|\Phi^{[1]}_{i} \rangle \\ \nonumber
        &- \sum_{i,j=1}^{N} \bigl|\langle \Phi^{[1]}_{i} |\,U_{\rm HVA}^\dagger(\theta)\,H\,U_{\rm HVA}(\theta)|\Phi^{[1]}_{j}\rangle\bigr|^{2}\, ,
\end{align}
where the $|\Phi^{[1]}_i\rangle$ denote the undressed 1QP basis states and $N$ is the number of sites in the cluster. To see why this cost function achieves block-diagonalization, let us, by analogy to Eq.~\eqref{eq:superposed_1qp}, also denote the prepared 1QP states by $|\chi_i\rangle = U_{\rm HVA}|\Phi^{[1]}_i\rangle$; they are also superpositions of $\ket{\Psi^{[1]}_{i}}$. Since $U_{\rm HVA}$ is unitary, we can extend $\{|\chi_i\rangle\}_{i=1}^{N}$ to a complete orthonormal basis $\{|\chi_\alpha\rangle\}$ of the full Hilbert space. Inserting the resolution of identity $\mathds{1} = \sum_\alpha |\chi_\alpha\rangle\langle\chi_\alpha|$ into the first term of Eq.~\eqref{eq:cf_variance},

\begin{align}
    \sum_{i=1}^{N} \langle\chi_i|H^2|\chi_i\rangle &= \sum_{i=1}^{N} \langle\chi_i|H \left(\sum_\alpha |\chi_\alpha\rangle\langle\chi_\alpha|\right) H|\chi_i\rangle \notag \\
    &= \sum_{i=1}^{N} \sum_{\alpha} \bigl|\langle\chi_\alpha|H|\chi_i\rangle\bigr|^2 \notag \\
    &= \sum_{i=1}^{N} \sum_{\alpha \in \mathrm{1QP}} \bigl|\langle\chi_\alpha|H|\chi_i\rangle\bigr|^2 \nonumber\\
    &\quad \quad \quad+ \sum_{i=1}^{N} \sum_{\alpha \notin \mathrm{1QP}} \bigl|\langle\chi_\alpha|H|\chi_i\rangle\bigr|^2.
    \label{eq:variance_expand}
\end{align}

The first sum on the right-hand side is exactly the second term in Eq.~\eqref{eq:cf_variance}, since $\alpha \in \mathrm{1QP}$ runs over $j = 1, \ldots, N$. Substituting back, the two terms cancel and the cost function reduces to

\begin{equation}
    C_\mathrm{var}^\mathrm{1QP}(\boldsymbol{\theta}) = \sum_{i=1}^{N} \sum_{\alpha \notin \mathrm{1QP}} \bigl|\langle\chi_\alpha|H|\chi_i\rangle\bigr|^2.
    \label{eq:variance_offdiag}
\end{equation}

This sum vanishes if and only if every matrix element of $H$ connecting the transformed 1QP sector to the rest of the Hilbert space is zero, i.e., the transformed 1QP sector is an invariant subspace of $H$, meaning that $H$ maps this subspace onto itself. Since both terms in Eq.~\eqref{eq:cf_variance} depend only on the subspace spanned by $\{|\chi_i\rangle\}$ and not on the choice of basis within it, $C_\mathrm{var}^\mathrm{1QP}$ is invariant under rotations within the 1QP block~\cite{sumeet2025}. 

Furthermore, the ground-state energy $E^{[0]}$, as shown in Eq.~\eqref{eq:1qp_hamiltonian}, must also be measured. Either the same adiabatic sweep as mentioned before can be used for this purpose, or another VQE trained using a standard energy-minimizing cost function~\cite{peruzzo_variational_2014} given as, 
\begin{align}\label{cf:energy}
    C_E(\theta) = \bra{\Phi^{[0]}}U_{\rm HVA}^\dagger(\theta) H U_{\rm HVA}(\theta) \ket{\Phi^{[0]}},
\end{align}
can be used to obtain the ground-state energy. In this work, we compute the ground-state energy using sample-based quantum diagonalization (SQD), as explained in Sec.~\ref{sec:subspace_diag}.

\subsection{Non-linear processing on clusters}\label{subsec:PCAT}

An exact block-diagonalizing unitary is not unique: for degenerate subspaces, infinitely many unitaries achieve the same decoupling. In practice, the unitary $U_{\mathrm{QA}}$ block-diagonalizes $H$ only approximately but the resulting effective Hamiltonian is not guaranteed to be cluster-additive, a property required for the NLCE to converge. In this section, we describe the PCAT~\cite{Hormann2023a} that enforces cluster additivity, and detail how its required matrix elements are obtained from the quantum device.

\subsubsection{Projective cluster-additive transformation}

 In an NLCE, thermodynamic-limit quantities are reconstructed from calculations on small connected clusters via a procedure that converges only if the effective Hamiltonian $H_\mathrm{eff}$ on two disconnected clusters $A$ and $B$ decomposes as

\begin{equation}
    H_\mathrm{eff}(A \cup B)
    = H_\mathrm{eff}(A) \otimes \mathds{1}_B
    + \mathds{1}_A \otimes H_\mathrm{eff}(B).
    \label{eq:cluster_additivity}
\end{equation}

Without such cluster additivity, the effective Hamiltonian can contain spurious terms that allow dressed quasi-particles to hop between disconnected clusters, causing the NLCE to diverge with increasing cluster size. Standard block-diagonalization transformations like Schrieffer-Wolff~\cite{bravyi2011} can violate this property, which motivates the process described below.

The PCAT~\cite{Hormann2023a} resolves this by constructing a modified unitary $U_{\text{PCAT}}$ from the QA solution, which preserves Eq.~\eqref{eq:cluster_additivity} by construction. The key insight is that eigenstates on disconnected clusters $A \cup B$ have product structure,
\begin{equation}
    |\Psi_{A \cup B}^{[n+m]}\rangle = |\Psi_A^{[n]}\rangle \otimes |\Psi_B^{[m]}\rangle.
    \label{eq:product_structure}
\end{equation}
For ground states on disjoint clusters, this product structure is automatically preserved by any block-diagonalizing transformation, since ground-state additivity, 
\begin{equation}
    E_{A\cup B}^{[0]}= E_A^{[0]} + E_B^{[0]},
    \end{equation}
always holds for disconnected clusters, which follows from the factorization of the Hamiltonian in Eq.~\eqref{eq:cluster_additivity}. For excited states, however, general unitary transformations $U_{\rm QA}$ mix in ground-state components that break the product structure in Eq.~\eqref{eq:product_structure}. PCAT eliminates this mixing by constructing modified states $|\tilde{\Psi}_i^{[1]}\rangle$ by eliminating projections onto the 0QP sector of $H_0$,
\begin{equation}
|\tilde{\Psi}_i^{[1]}\rangle = |\Psi_i^{[1]}\rangle - \frac{\langle \Phi^{[0]} | \Psi_i^{[1]} \rangle}{\langle \Phi^{[0]} | \Psi^{[0]} \rangle} |\Psi^{[0]}\rangle
    \label{eq:modified_states}
\end{equation}

where $|\Psi^{[0]}\rangle$ and $|\Psi^{[1]}_i\rangle$ are the exact ground-state and 1QP eigenstates of $H$, and $|\Phi^{[0]}\rangle$ is the undressed ground state. Note that Eq.~\eqref{eq:modified_states} implicitly assumes $\langle \Phi^{[0]}|\Psi^{[0]}\rangle \neq 0$, which is guaranteed by the adiabatic connection between the dressed and undressed ground states. 

As the states $|\tilde{\Psi}^{[1]}_i\rangle$ have zero projection onto the ground-state sector of $H_0$, i.e., $P_0|\tilde{\Psi}^{[1]}_i\rangle = 0$, the product structure on disconnected clusters is restored, \mbox{$|\tilde{\Psi}^{n+m}_{A\cup B}\rangle = |\tilde{\Psi}^n_A\rangle \otimes |\tilde{\Psi}^m_B\rangle$}, and the overlap matrix between undressed and modified states becomes block-diagonal in the cluster indices.
We note that, among the models studied in this work, the modification in Eq.~\eqref{eq:modified_states} is only needed for the TFIM with a longitudinal field (TFIM+LF). For the remaining models, the $\mathbb{Z}_2$ symmetry enforces $\langle \Phi^{[0]}|\Psi^{[1]}_i\rangle = 0$, so that $|\tilde{\Psi}^{[1]}_i\rangle = |\Psi^{[1]}_i\rangle$.

The states $|\Psi^{[0]}\rangle$ and $|\Psi^{[1]}_i\rangle$ do not need to be measured directly on the quantum device, but only in expectation values and overlap matrices as we describe in Sec.~\ref{sec:measurements}.  Thus, from the modified states $|\tilde{\Psi}^{[1]}_i\rangle$, we construct the modified overlap matrix~\cite{Hormann2023a}
\begin{equation} 
    \tilde{O}_{ij}^{[1]} = \langle \Phi_i^{[1]} | \tilde{\Psi}_j^{[1]} \rangle,
   \label{eq:modified_overlap}
\end{equation}
and the PCAT correction within the 1QP sector,
\begin{equation}\label{eq:pcat_correction}
    V^{[1]} = \tilde{O}^{[1]\dagger} \left( \tilde{O}^{[1]} \tilde{O}^{[1]\dagger} \right)^{-1/2}.
\end{equation}
Eq.~\eqref{eq:pcat_correction} is the symmetric Löwdin orthonormalization of $\tilde{O}^{[1]\dagger}$, the partial isometry closest to the identity that orthonormalizes the modified states \cite{Hormann2023a}. The effective 1QP Hamiltonian is then
\begin{equation}
    H_{\text{eff}}^{[1]} = V^{[1]\dagger} H^{[1]} V^{[1]} - E^{[0]},
    \label{eq:eff_ham_PCAT}
\end{equation}
where $H^{[1]}_{ij} = \langle\Psi^{[1]}_i|H|\Psi^{[1]}_j\rangle$. By construction, $V^{[1]}$ inherits the block-diagonal structure of $\tilde{O}^{[1]}$ on disconnected clusters, guaranteeing that the resulting $H^{[1]}_\mathrm{eff}$ satisfies Eq.~\eqref{eq:cluster_additivity}.
We note that the matrix inversion and square root in Eq.~\eqref{eq:pcat_correction} are the non-linear operations that we mentioned earlier.

\subsubsection{Computing PCAT on quantum hardware}

The final result $V^{[1]\dagger} H^{[1]} V^{[1]}$ in Eq.~\eqref{eq:eff_ham_PCAT} depends on $\tilde{O}^{[1]}$ only through the combinations $\tilde{O}^{[1]}\tilde{O}^{[1]\dagger}$ and $\tilde{O}^{[1]}\tilde{H}^{[1]}\tilde{O}^{[1]\dagger}$, which are unchanged if one rotates the basis within the 1QP sector~\cite{Hormann2023a}. In practice, this means that the states prepared by the QA do not need to be eigenstates of $H$; it is sufficient that they span the correct subspace. We can therefore use the QA-prepared states $|\chi_i\rangle$ directly in Eqs.~(\ref{eq:modified_states})-(\ref{eq:eff_ham_PCAT}) without requiring any additional orthonormalization step.

The corresponding implementation of Eq.~\eqref{eq:modified_states} reads
\begin{equation}
    |\tilde{\chi}^{[1]}_i\rangle = |\chi^{[1]}_i\rangle - \frac{\langle \Phi^{[0]}|\chi^{[1]}_i\rangle}{\langle \Phi^{[0]}|\chi^{[0]}\rangle} |\chi^{[0]}\rangle,
    \label{eq:modified_chi}
\end{equation}
where $|\chi^{[0]}\rangle$ and $|\chi^{[1]}_i\rangle$ are the ground-state and 1QP states prepared by the quantum algorithm. This substitution is justified by the subspace invariance: the PCAT result depends only on the subspace spanned by the prepared states, not on whether they are exact eigenstates.

The PCAT construction in Eq.~\eqref{eq:eff_ham_PCAT} requires two sets of measurements from the quantum device, each forming an $(N+1) \times (N+1)$ matrix for a cluster of $N$ sites. 

The ground-state energy is computed independently (see Sec.~\ref{sec:subspace_diag}). Therefore, in order to build the PCAT construction in Eq.~\eqref{eq:eff_ham_PCAT}, we only need to measure the ground-state energy and two $N \times N$ matrices corresponding to the 1QP sector. These two matrices are as follows.

\begin{enumerate}
\item \textbf{Hamiltonian matrix $H$}: Here, the elements $H_{ij} = \bra{\Phi_i^{[1]}}  U^\dagger_{\rm QA} H U_{\rm QA}  \ket{\Phi_j^{[1]}}$ $=\bra{\chi_i^{[1]}}  H   \ket{\chi_j^{[1]}}$ are in the QA-prepared state basis. The diagonal elements are standard expectation values, whereas off-diagonal elements require measurements on pairwise superpositions of computational basis states, as we explain in the next section.

\item \textbf{Overlap matrix $\tilde{O}$}: Elements $\tilde{O}_{ij} = \langle \Phi^{[1]}_i | \tilde{\chi}_j^{[1]} \rangle$ quantify the overlap between undressed eigenstates $|\Phi^{[1]}_i\rangle$ and modified QA-prepared states $\ket{\tilde{\chi}^{[1]}_i}$. These are constructed by preparing the QA-prepared states $\ket{\chi^{[1]}_i}$, and applying the correction in Eq.~\eqref{eq:modified_chi} if it is needed. The correction is computed separately as described in Sec.~\ref{sec:measurelf}. Both can be measured using the Hadamard or the CX-test described in Sec.~\ref{subsec:overlapmat}. 

\end{enumerate}

The total measurement cost scales as $\mathcal{O}(N^2)$, which is the number of matrix elements per cluster. In the case of VQE, these measurements are performed only after optimization is complete, and the subsequent PCAT construction and NLCE embedding proceed entirely classically, since all matrices are of quadratic size with respect to the number of spins $N$ in the Hamiltonian.\\ 

\subsection{Measuring matrix elements on a quantum device}\label{sec:measurements}

We have described three sorts of measurements that we require from the quantum device; each needs its own treatment. 

\subsubsection{Ground-state energy}\label{sec:subspace_diag}

For the calculation of the ground-state energy, we train the VQE on a classical computer, with an energy minimization cost function, as in Eq.~\eqref{cf:energy}. For ASP, which we use as well, no training is required. On the quantum device, we then sample the state obtained with the quantum circuit in the $Z$-basis, and perform sample-based quantum diagonalization  (SQD)~\cite{kanno_QuantumSelected_2023,stockinger2026}, which consists of building the sub-matrix of the Hamiltonian in the subspace formed by the sampled computational basis states, and classically diagonalizing that matrix. This works particularly well when the ground state consists of a superposition of a polynomial number of computational basis states. Hence, assuming that the ground state $\ket{E_0}$ of some $N$-qubit Hamiltonian $\hat{H}$ can be well approximated by a linear combination of $M$ computational basis states,
and that all these have been sampled in the data acquisition phase, we need to diagonalize an $M \times M$ matrix.

By the variational principle, its lowest eigenvalue upper bounds the ground-state energy of $\hat{H}$, and the bound tightens as the support of the ground state on the sampled bitstrings increases.

\subsubsection{Hamiltonian matrix}

Beyond the ground-state energy, we must measure off-diagonal Hamiltonian elements $H_{ij} = \bra{\Phi_i^{[1]}}  U^\dagger H U  \ket{\Phi_j^{[1]}}$ for $i,j = 1,...,N$. Here $U$ denotes any unitary acting on the full $2^N$ Hilbert space, either VQE or ASP. Recall that $\ket{\Phi_j^{[1]}}$ are Hamming weight one computational basis states, $ \ket{\Phi_j^{[1]}}=X_j\ket{0}^{\otimes N} $, where $X_j$ is the Pauli $X$ matrix acting on the $j$th qubit. In the general case of any Hamiltonian and any pair of states whose superposition we can prepare, it is possible to extract two off-diagonal elements from four expectation values. Consider the expectation values
\begin{align*}
    \gamma_1 &= \frac12 (\bra{\Phi^{[1]}_i} + \bra{\Phi^{[1]}_j}) U^\dagger H U (\ket{\Phi^{[1]}_i} + \ket{\Phi^{[1]}_j}) \\
    \gamma_2 &= \frac12 (\bra{\Phi^{[1]}_i} - \bra{\Phi^{[1]}_j}) U^\dagger H U (\ket{\Phi^{[1]}_i} - \ket{\Phi^{[1]}_j}) \\
    \gamma_3 &= \frac12 (\bra{\Phi^{[1]}_i} - {\rm i}\bra{\Phi^{[1]}_j}) U^\dagger H U (\ket{\Phi^{[1]}_i} + {\rm i}\ket{\Phi^{[1]}_j}) \\
    \gamma_4 &= \frac12 (\bra{\Phi^{[1]}_i} + {\rm i}\bra{\Phi^{[1]}_j}) U^\dagger H U (\ket{\Phi^{[1]}_i} -{\rm i} \ket{\Phi^{[1]}_j}) 
    \end{align*}
where we recall that $(\ket{\Phi^{[1]}_i} -{\rm i} \ket{\Phi^{[1]}_j})^\dag = \bra{\Phi^{[1]}_i} + {\rm i}\bra{\Phi^{[1]}_j}$. Then 
\begin{align*} \label{eq:hgammaelem}
    \gamma_1 - \gamma_2 &= 2 \mathrm{Re}\bra{\Phi^{[1]}_i}U^\dagger H U\ket{\Phi^{[1]}_j} \\ 
    -{\rm i}(\gamma_3 - \gamma_4) &= 2\mathrm{Im}\bra{\Phi^{[1]}_i}U^\dagger H U\ket{\Phi^{[1]}_j},
\end{align*}
Due to the Hermiticity of the matrix, only half of the off-diagonal elements need to be computed. 

The superposition $(\ket{\Phi^{[1]}_i} + \ket{\Phi^{[1]}_j})/\sqrt{2}$ can be built with $X_i \,{\mathrm{CNOT}}_{ij} \,H_i \ket{0}^{\otimes N} $, and likewise for the three other superpositions.

\subsubsection{Overlap matrix}\label{subsec:overlapmat}

Finally, we must compute overlaps between states, or, in other words, off-diagonal elements of a unitary, in order to build the \textit{overlap matrix} $\tilde{O}_{ij} = \braket{\Phi^{[1]}_i | \tilde{\chi}^{[1]}_j}$, where $\ket{\tilde{\chi}^{[1]}_j}$ is the QA-prepared superposition as defined in Eq.~\eqref{eq:modified_chi} with the PCAT correction applied. 
When $\langle \Phi^{[0]}|\chi^{[1]}_i\rangle=0$, the elements we compute reduce to $\braket{\Phi^{[1]}_i|\chi^{[1]}_j} = \bra{\Phi^{[1]}_i}U_{\rm QA}\ket{\Phi^{[1]}_j}$.

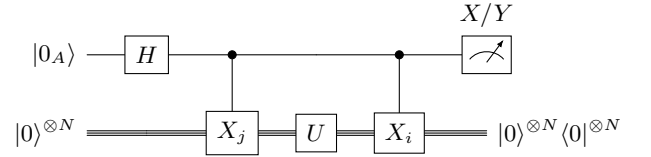
\begin{figure}[h!]
    \centering
        \begin{quantikz}[wire types={q,b}, thin lines=true]
        \lstick{$\ket{0_A}$} &\gate{H}&\ctrl{1}   &   & \ctrl{1} & \meter{X/Y} \\
        \lstick{$\ket{0}^{\otimes N}$} &     &\gate{X_j} &  \gate{U} & \gate{X_i}  & \rstick{$\ket{0}^{\otimes N}\bra{0}^{\otimes N}$} 
        \end{quantikz}
    \caption{The CX-test. In the second register, the controlled $X$ gates indices indicate which qubit it is applied to, out of the $N$ qubits.}
    \label{fig:steftest}
\end{figure}

For any unitary $U$, this can usually be done with the well-known Hadamard test, which we summarize for completeness in App.~\ref{app:hadamtest}. The main disadvantage of this test is that it requires us to implement a controlled version of $U$, and this might make the circuit very deep. For PCAT, however, the controlled unitary can be avoided entirely, reducing the cost of controlled unitaries to two CX gates per matrix element.  Consider the circuit in Fig.~\ref{fig:steftest}.

In that circuit, we select for the all-zeros state in the system register, and measure the ancilla as in the Hadamard test. In what follows, $\ket{0}^{\otimes N}\equiv \ket{0}$ for clarity. 
The expectation of the composite state of the ancilla and system is
\begin{align*}
 \bra{\mu'} (X \otimes \ket{0}\bra{0}) \ket{\mu'} &= \frac{1}{2} \left[ \bra{0}U^\dagger\ket{0}\bra{0}X_i U X_j\ket{0} + h.c. \right] \\ 
 & =  \mathrm{Re}\left(\bra{0}U^\dagger \ket{0} \bra{\Phi^{[1]}_i}U\ket{\Phi^{[1]}_j}\right), 
\end{align*}
where we defined
\begin{align*}
 \ket{\mu'} &= \frac{1}{\sqrt{2}} \left[ \ket{0_A} U \ket{0} + \ket{1_A}X_i U X_j\ket{0} \right],
\end{align*}
to be the output state of the circuit before the measurement. We immediately notice that this circuit gives us the desired matrix element multiplied by a scalar $\bra{0}U^\dagger\ket{0}$. This means we only need to measure this scalar once, using the Hadamard test, and the rest of the $N^2$ values can be computed using the cheaper CX-test described above. This is particularly useful when a polynomial amount of matrix elements need to be computed. It can also be straightforwardly extended to the imaginary part of the matrix element.

As we mentioned above, in the specific case of PCAT, we can also avoid computing $\bra{0}U^\dagger\ket{0}$ entirely, allowing us to avoid any use of the Hadamard test. To see this, consider again the PCAT correction defined in Eq.~\eqref{eq:pcat_correction}, and the resulting effective Hamiltonian. Assuming we use the resulting matrix directly from the CX-test, i.e. multiplied by $\bra{0}U^\dagger\ket{0}$, we get
\begin{align}
     \sqrt{\frac{\bra{0}U\ket{0}}{\bra{0}U^\dagger\ket{0}}} V^{[1]}\, ,
\end{align}
and the resulting prefactors cancel in Eq.~\eqref{eq:eff_ham_PCAT}. Therefore, we do not need to compute $\bra{0}U^\dagger\ket{0}$ at all. Hence, we can entirely skip the expensive Hadamard tests, and simply perform the CX-tests. Still, we will see that, even with this improvement, ASP on current hardware does not produce results as accurate as those produced by the VQE.

We note the similarity between our CX-test and the iterative Hadamard test and related algorithms independently discovered by \textcite{schiffer_Hardwareefficient_2025}. While their approach also reduces the number of controlled gates needed for overlap estimation, the CX-test is more efficient in the PCAT setting because the global prefactor $\langle 0|U^\dagger|0\rangle$ cancels in the final product $V^{[1]\dagger}H^{[1]} V^{[1]}$, eliminating the need to measure it altogether.

Apart from the above mentioned measurements, we note that VQE training itself requires evaluating the cost function Eq.~\eqref{eq:cf_variance} on hardware, adding to the total measurement budget. In this work, however, all VQE optimizations are performed on a classical computer, therefore we measure only the final matrix elements on the QPU. Our goal is to demonstrate the possibility of performing the NLCE+QA set up on a quantum device which includes the non-linear PCAT. As quantum hardware improves, it will become possible to replace the VQE with ASP, which performs the same block-diagonalizing function while forgoing the training process and sidestepping associated challenges such as barren plateaus~\cite{larocca_Review_2024}. 

\subsection{Treatment of mixing between 0QP and 1QP}\label{sec:measurelf}
When the overlap $\braket{\Phi^{[0]} | \chi^{[1]}_j}$  is non-zero,  the second term in Eq.~\eqref{eq:modified_chi} does not vanish. Hence, elements of $\tilde{O}_{ij}$ cannot be simply computed using $\braket{\Phi^{[1]}_i|U|\Phi^{[1]}_j}$, but we must now take the second term of Eq.~\eqref{eq:modified_chi} into account. Substituting $|\chi^{[n]}_j\rangle =U|\Phi^{[n]}_j\rangle, n \in \{0,1\}$ and multiplying Eq.~\eqref{eq:modified_chi} by $\langle \Phi^{[1]}_i|$, each matrix element is then computed as
\begin{align*}
  \tilde{O}_{ij} & = \braket{\Phi^{[1]}_i|U|\Phi^{[1]}_j} - \frac{\langle \Phi^{[0]} | U | \Phi_j^{[1]} \rangle}{\langle \Phi^{[0]} | U|\Phi^{[0]} \rangle} \braket{\Phi^{[1]}_i|U |\Phi^{[0]}} \\
  & = \braket{\Phi^{[1]}_i|U|\Phi^{[1]}_j} - \Delta_{ij}.
\end{align*}
Beyond the $N^2$ terms needed to compute $\braket{\Phi^{[1]}_i|U|\Phi^{[1]}_j}$, as we described earlier, we also need to compute three overlaps in the correction that is subtracted from the main term. These can be computed on their own, each with a single CX-test. The phase cancellation, described in Sec.~\ref{subsec:overlapmat}, is still valid, and the Hadamard test for $\braket{0|U|0}$ can still be avoided, but one must be careful with the unitaries. While the ASP algorithm identically prepares superpositions of 1QP states as it prepares the ground state, the unitaries in the above equation may differ when using VQEs: the circuit preparing the 1QP states may be different from the one preparing the ground state. This can be circumvented by adding a ground state term to the block-diagonalizing cost function in Eq.~\eqref{eq:variance_offdiag} that also minimizes the ground-state energy, such that the CX-test phases of the denominator and numerator cancel, and the overall phase is the same as the first term~\cite{sumeet2025}.

In total, the correction terms $\Delta_{ij}$ require $2(2N+1)$ more circuits, because their product can be computed independently. Regardless, it turns out that it is impossible to detect this correction $\Delta_{ij}$ within our current sampling budget $M$, i.e. $\Delta_{ij} \ll 1/\sqrt{M}$. Indeed, for the studied TFIM+LF case, we found that $\Delta_{ij} \approx 4 \cdot 10^{-4}$, thus requiring about $M= 5 \cdot 10^6$ shots per element.

\subsection{Numerical linked-cluster expansions}
\label{subsec:nlce}

The preceding subsections described how to obtain the
effective 1QP Hamiltonian on a single finite cluster $C$.
We now focus on how these finite-cluster results are embedded in the thermodynamic limit within numerical linked-cluster expansions (NLCEs), which provide a systematic scheme for obtaining bulk properties from finite-cluster calculations. The core idea behind NLCEs is to decompose an extensive quantity into contributions from connected clusters using the inclusion-exclusion principle. Reduced contributions are computed on topologically distinct clusters and embedded into the infinite lattice, weighted by the number of embeddings per site.

For a property $P$, the reduced contribution of a cluster $C$ is defined recursively as
\begin{equation}
    \bar{P}_C = P_C - \sum_{C' \subsetneq C} w_{C'/C} \, \bar{P}_{C'},
    \label{eq:reduced_contribution}
\end{equation}
where $w_{C'/C}$ counts embeddings of subcluster $C'$ within $C$. The thermodynamic-limit value is then obtained by summing reduced contributions over all topologically distinct clusters,
\begin{equation}
    P = \sum_C w_C \, \bar{P}_C,
    \label{eq:nlce_sum}
\end{equation}
where $w_C$ is the embedding weight of cluster $C$ per lattice site.

For computational efficiency, it is optimal to employ a rectangular graph expansion in which only clusters of dimensions $L_m \times L_n$ with $L_m L_n \leq N_{\text{max}}$ are included. Here $N_{\text{max}}$ is the number of spins in the largest cluster taken in the NLCE calculation. This restricts the number of clusters to polynomial growth in system size, making the approach tractable for hybrid quantum-classical implementation \cite{sumeet2025}. In particular, for one-dimensional lattices like the chain and the two-leg ladder investigated in this work, the number of clusters reduces to the two:
$N_{\text{max}}\times 1$ and $(N_{\text{max}}-1)\times 1$ for the chain  and $(N_{\text{max}}/2)\times 2$ and $(N_{\text{max}}-2)/2\times 2$ for the two-leg ladder.

A critical requirement for NLCE convergence is that the quantity $P$ be cluster-additive: for any two disconnected clusters $A$ and $B$, the reduced contribution of their union must vanish, $\bar{P}_{A \cup B} = 0$. This property ensures that only connected clusters contribute to the expansion which is guaranteed by PCAT, as explained in section~\ref{subsec:PCAT}.

For each finite cluster $C$, the QA and PCAT procedure described above yields an effective 1QP Hamiltonian
$H^{[1]}_{\mathrm{eff}}$ with real-space matrix elements
connecting sites within the cluster. Assuming a one-site unit cell, the 1QP dispersion $\omega(\vec{k})$
in the thermodynamic limit is then obtained from the NLCE embedding through a Fourier transform,

\begin{equation}
    \omega(\vec{k}) = \sum_{L_m, L_n}
    \sum_{\vec{\mu}, \vec{\nu}}
    e^{i\vec{k} \cdot (\vec{\nu} - \vec{\mu})}\,
    \bar{H}^{[1]}_{\mathrm{eff}, L_m \times L_n,
    \vec{\mu}, \vec{\nu}},
    \label{eq:dispersion}
\end{equation}

where $\bar{H}^{[1]}_{\mathrm{eff}}$ denotes the reduced
contribution of each cluster, obtained according to Eq.~\eqref{eq:reduced_contribution}, and the outer sum runs
over all clusters $L_m \times L_n$ with weight one included in the NLCE. For a larger unit cell with $m$ sites, one obtains by the same embedding procedure an $m\times m$ matrix for each momentum $\vec{k}$ resulting in $m$ 1QP bands.

\section{Error and uncertainty}\label{sec:error_analysis}

Because the shot noise, the hardware imperfections, and the non-linear character of the post-processing interact in a nuanced way, it is important to carefully build an understanding of how the uncertainty and the error propagate into the final dispersion obtained by NLCE+QA in the thermodynamic limit. We use the word ``uncertainty" to denote the lack of knowledge due to the inevitable need to estimate observables with a finite number of measurements, and ``error" to refer to the difference between the ideal value and the imperfect estimate. 

\subsection{Uncertainty propagation}

We first discuss how the final uncertainty in the dispersion relations is calculated, or, in other words, how we obtain statistical uncertainty through the non-linear post-processing. The non-linear character of the PCAT construction, particularly the matrix inverse and square root in $( \tilde{O}^{[1]} \tilde{O}^{[1]\dagger})^{-1/2}$, means that errors propagate in a non-trivial manner and can be amplified significantly if the input data is not sufficiently accurate. Computing the propagation of uncertainty from the initial quantum measurements to the final dispersion relation requires the use of a Monte Carlo simulation~\cite{zhang_Modern_2020}. From the expectation values measured on the quantum device, we compute the error-on-the-mean (EOTM)~\cite{GUM_2008,arute_Quantum_2019}, which is defined for each Pauli string $P$ as $\sigma = \sqrt{(1 - \langle P \rangle^2)/M}$, with $M$ the number of shots. It gives the best estimates for the variance of the mean value, and is also called the experimental standard deviation of the mean~\cite{GUM_2008}. 

To analyze the error propagation through our post-processing procedure, we generate random data points (overlap and Hamiltonian matrices) that are distributed according to a Gaussian distribution around the found mean value with variance $\sigma^2$. We then propagate these data points through the NLCE procedure to compute dispersion curve values $\omega(\vec{k})$ for each of them. This allows us to take the standard deviation of the resulting distribution of $\omega(\vec{k})$-values as uncertainties for our results

We sample $M_{\rm mc}=10^4$ dispersions, based on a study of EOTM convergence shown in App.~\ref{app:mcconv}.

As matrix sizes increase and Monte Carlo sampling becomes too difficult, one could turn to methods such as the Unscented Transform (UT)~\cite{wan_unscented_2000} to compute the uncertainty propagation at a lower cost. In our case, this would decrease the number of samples needed from about $10^4$ to $\mathcal{O}(N^2)$, where $N$ is the number of qubits. However, the UT has stricter assumptions on the form of the distribution, and may diverge if the function is not smooth enough~\cite{wan_unscented_2000}.

For the ground-state energy estimation we use a conservative upper bound $h/\sqrt{M}$, with $h$ as defined in Eq.~\eqref{eq:ham_tfim}, since the uncertainty in SQD is relative to the distribution of the ground state~\cite{kanno_QuantumSelected_2023}, and not straightforwardly computable as it involves diagonalizing a matrix. Yet, simulations indicate that the value is likely lower. 

\subsection{Error propagation}

Having discussed the uncertainty of our results due to shot noise, we now turn to investigate the influence of gate errors on our results. To analyze these effects we do noisy simulations for different noise levels, and explore their influence on the obtained dispersions. This allows us to investigate which noise level leads to comparable results to those that we obtain from our runs on the real QPU, and helps us determine which number of measurements is needed to determine the dispersion values with good enough accuracy.

Our noisy simulations use the assumption that depolarization noise is the main hardware error channel~\cite{cai_Quantum_2023,urbanek_Mitigating_2021}. This is justified when the circuit has many qubits and gates, as the different errors can be amalgamated into depolarization noise~\cite{nielsen_Quantum_2010}. While one may of course build more advanced error models, which, for example, take non-unital noise into account~\cite{mele_Noiseinduced_2026}, comparisons with real hardware runs indicate that it is sufficiently justified in terms of experimental performance.

Depolarization noise can be expressed as a quantum channel $\mathcal{E}$ for an $N$-qubit state $\rho$, written as 
\begin{align}\label{eq:depolchaneq}
    \mathcal{E}(\rho) = (1-p)\,\rho + p \,\,\mathrm{tr}(\rho) \, \frac{\mathds{1}}{2^N} ,
\end{align}
where $p$ is the depolarization rate and $\mathds{1}/2^N$ describes the maximally mixed state.
We analyze the effect of depolarization noise by applying the depolarization channel as given in Eq.~\eqref{eq:depolchaneq} after each gate. The effect of this gate-based approach will result in an effective depolarization rate for observables that depends on both the circuit and the observable.

As the emulator for AQT devices uses depolarization rates of $p=0.003$ for single-qubit gates, and $p=0.01$ for two-qubit gates, we take these values as reference (though we note that they are not finetuned to exactly match the particular device we use). In our numerical simulations, we can then increase or decrease these depolarization rates to see how they distort the dispersion relation.

\begin{figure*}[t!]
    \centering
    \includegraphics[width=.75\linewidth]{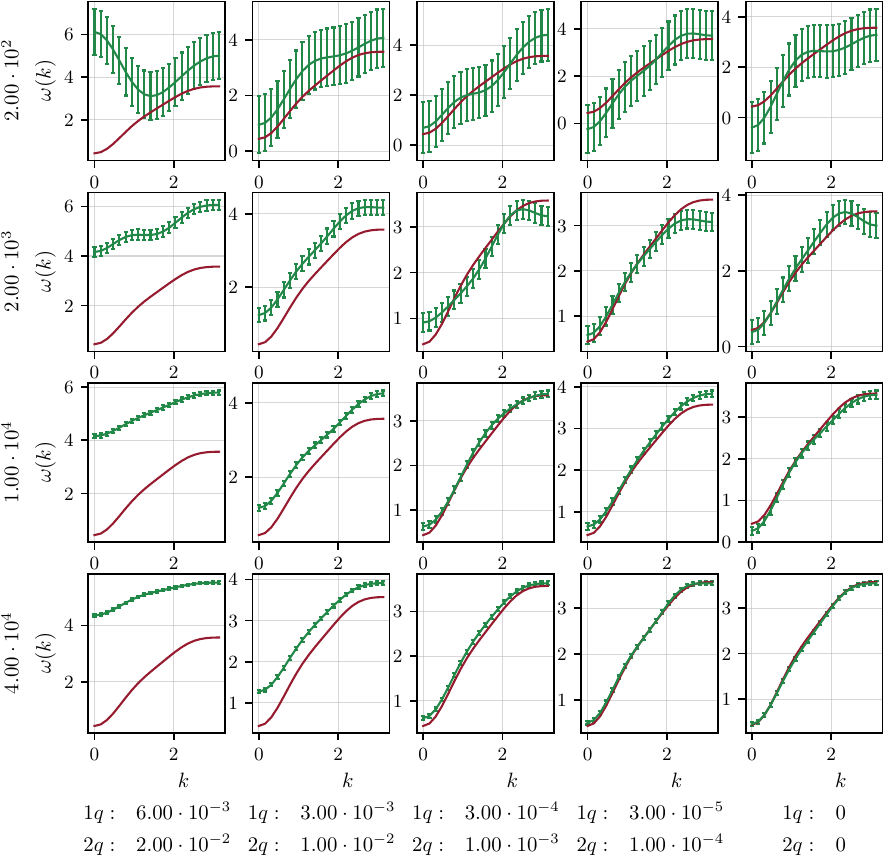}
    \caption{Interplay between shot noise and depolarization noise when computing the dispersion curve, using noisy simulation, for a TFIM chain at $J/h=0.8$ obtained by NLCE with $N_{\rm max}=5$. The $Y$-axis label shows the number of shots, while the $X$-axis label shows the one-qubit and the two-qubit  gate depolarization rates, indicated by $1q$ and $2q$. The red line represents the noiseless statevector simulation, and the green line a noisy simulation. The error bars denote one standard deviation of statistical uncertainty.}
    \label{fig:depol_study}
\end{figure*}

In Fig.~\ref{fig:depol_study} we investigate the effect of various levels of depolarization noise on the dispersion that we compute via the non-linear post-processing of the noisy measured data. 
One can see a tendency of the dispersion curve computed from the data to increasingly shift upwards as the level of depolarization noise increases. This trend becomes more pronounced as we increase the number of shots to reduce the uncertainty in the data.

In our case, the upward bias originates from the way we measure the ground energy using SQD, which gives more accurate results than the other measurements. We obtain the  dispersion via Eq.~\eqref{eq:dispersion}, with effective Hamiltonians given by Eq.~\eqref{eq:eff_ham_PCAT}. Now, for a depolarizing channel at the gate level as in Eq.~\eqref{eq:depolchaneq}, one can obtain an effective depolarization channel for each observable $O$ on an $N$-qubit state $\rho$ as
\begin{align} \label{eq:depolchanobs}
    \braket{O_{\mathrm{noisy}}}_\rho &= (1-p_O)\braket{O_{\mathrm{ideal}}}_\rho + p_O \,\,\mathrm{tr}\big(\frac{O}{2^N}\big),
\end{align}
where $p_O$ is the effective depolarization rate of the observable $O$ for $\rho$. In our case, we can assume that the matrix elements of $H_{ij}$ have a common effective depolarization rate, which we call $p_H$, while Eq.~\eqref{eq:pcat_correction} tells us that any depolarization noise in the overlap matrix cancels. The  
SQD technique allows us to get a much more accurate energy estimate than direct estimation. We can thus assume an effective depolarization rate $p_E$ for the ground state energy that is smaller than $p_H$, and write $p_H = p_E + \delta$ with $\delta >0$.

Hence, the depolarized effective Hamiltonian can be taken to read 

\begin{align*}
H^{[1]}_{\mathrm{eff,noisy}} &= (1-p_H) V^{[1]\dagger} H^{[1]} V^{[1]} - (1-p_E) E^{[0]} \\
&=  (1-p_H) H^{[1]}_{\mathrm{eff}} - \delta E^{[0]},
\end{align*}
where we used $p_E = p_H - \delta$. In the resulting dispersion given by  Eq.~\eqref{eq:dispersion}, we can take the constant $-\delta E^{[0]}$ contribution out of the sum, since it acts on the diagonal. As the ground state energy $E^{[0]}$ is always negative for the systems we study, the constant $-\delta E^{[0]}$ pushes the overall dispersion upwards. In turn, the $1-p_H$ factor, which contributes with various phase prefactors, compresses the vertical spread dispersion, as we see most illustratively in the bottom left panel of Fig.~\ref{fig:depol_study}. In fact, we can verify that our interpretation is correct by setting $p_E = p_H$; in this case, $\omega(k) \rightarrow 0, \, \forall k$ as $P_E = P_H \rightarrow 1$, and the bias $-\delta E^{[0]}$ disappears, as expected. Note that while this explains the upward bias at first order, the actual shape is, as we discussed, further distorted by shot noise, the fact that matrix elements do not actually have the same depolarization rate, and are themselves linear combinations of expectation values (see Eq.~\eqref{eq:hgammaelem}), as well as the fact that there are multiple clusters in the NLCE sum. 

\section{Results}\label{sec:results}

The simulations were programmed using qiskit~\cite{qiskit}, and the quantum experiments were performed on 5-8 qubits of an AQT Marmot 20-qubit trapped-ion quantum processing unit (QPU), with nominal single-qubit gate fidelity of $99.96\%$, and two-qubit gate fidelities of $98.47\%$~\cite{_MARMOT_}, seated at the Leibniz Supercomputing Centre (LRZ) in Garching, Germany. We examined three systems: the one-dimensional TFIM, the TFIM chain in a longitudinal field (TFIM+LF), and the two-dimensional TFIM ladder. 

For the cases involving a TFIM on a chain geometry, we performed one run for $J/h=0.3$, two runs for $J/h=0.8$ and three runs for $J/h=1$, as well as three runs for TFIM+LF on a chain for $J/h=0.5$ and $h_l=0.1$. For the TFIM on ladder, we performed a single run at $J/h=0.2$. Each run comprised 2000 shots per matrix element, and as much for the ground state. The number of runs and shot budget were determined by the available QPU access time. See Tab.~\ref{tab:tfimcases} for a summary.

For all cases combined, the total shot budget is about five million, which corresponds to around 430 hours of continuous runtime on the quantum device. Hence, due to the length of the calculations, which spanned several weeks, there is significant variability between the QPU runs. Despite this, the experimental dispersions generally adhere to the expectations, and more sampling could improve the stability of the results, as discussed in Sec.~\ref{sec:error_analysis}. We compare our hardware (QPU) and noisy simulation results either to an analytic solution (available only for the TFIM in 1D) or to exact diagonalization (ED) within the NLCE framework, which we refer to as NLCE+ED. We additionally compare to statevector (SV) simulations, isolating the error introduced by VQE training from that of hardware noise.

\subsection{Model}\label{sec:tfim}

The transverse-field Ising model (TFIM) is one of the paradigmatic models of the NISQ era, and a well-known test bench of quantum algorithms. The Hamiltonian for the TFIM in a longitudinal field is given by
\begin{equation}\label{eq:ham_tfim}
    H = -\, J\sum_{\langle \nu,\,\mu \rangle}X_\nu X_\mu - \,h \sum_{\nu} Z_\nu  - h_l \sum_{\nu} X_{\nu},
\end{equation}
where $X$ and $Z$ are Pauli matrices, $J$ denotes the Ising coupling for the nearest-neighbor spins, $h$ is the transverse field and $h_l$ is the longitudinal field. In the absence of this longitudinal field ($h_l=0$), the TFIM exhibits ferromagnetic order for $J > 0$ and antiferromagnetic order for $J < 0$ on bipartite lattices, in the limit $\abs{J} \gg h$. It possesses a global $\mathbb{Z}_2$ symmetry, corresponding to invariance under the simultaneous flip of all spins. The addition of the longitudinal field $h_l$ breaks the $\mathbb{Z}_2$ symmetry, thus leading to the mixing of the 0QP and 1QP states.

The model in Eq.~\eqref{eq:ham_tfim} is symmetric under reflection, such that for an $N$-qubit chain, $H_{N-j,N-k}=H_{j,k}$, and likewise for $\tilde{O}$. This also holds for the ladder geometry, which has a supplementary reflection symmetry about the centerline. Enforcing this symmetry allows us to effectively get twice the amount of shots per matrix element.

\subsection{Number of unique circuits}

\begin{figure}
    \centering
    \includegraphics[width=1\linewidth]{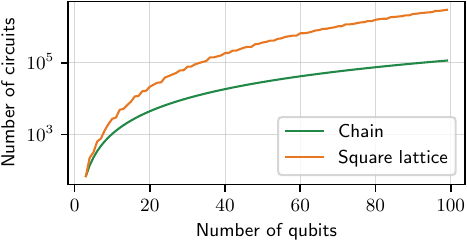}
    \caption{Number of unique circuits to compute the full NLCE+QA scheme, for the pure TFIM model on the chain and on the square lattice (using rectangular expansion for NLCE).}
    \label{fig:n_circ_needed}
\end{figure}

In Fig.~\ref{fig:n_circ_needed}, we show the number of unique circuits needed to compute an effective Hamiltonian of the TFIM model without longitudinal field for a one-dimensional chain and two-dimensional square lattice using the full NLCE+QA scheme, as a function of the number of qubits in the model.

By \textit{unique circuit}, we mean that, given the same algorithm, two different non-commuting Pauli string measurements will require different basis changes at the end, resulting in two distinct circuits. For the TFIM, the total is computed as follows: one circuit is needed for the ground state, if it is measured using SQD; otherwise, a normal energy measurement is required, as explained thereafter. Each overlap measurement requires two circuits, one for the real part, and the other for the imaginary part. There are therefore $\mathcal{O}(N^2)$ such circuits to compute for a given cluster $C$. For the Hamiltonian matrix measurements, each Pauli string measurement that is off-diagonal requires four circuits, while diagonal ones are real and require only one. One can obtain an energy expectation of the one-dimensional TFIM with two circuits, one for the single $Z$ terms, and one for the $XX$ terms. When adding a longitudinal field, we can use the $X$-basis measurements. Finally, in order to perform the NLCE, multiple lattice sizes must be computed~\cite{Hormann2023a}, and this depends both on the desired approximation level and dimensions. For a chain and a square lattice, the effective Hamiltonian for a single cluster of $N$ qubits requires
\begin{align*}
    n_{\mathrm{circ}}  &= 2N^2 + 2(2N^2-N) + 1 
\end{align*}
circuits. This number of unique circuits grows polynomially with system size, which remains a tractable number. Each circuit is easy to compute on quantum devices, while the square lattice becomes potentially difficult for classical ones. 

As mentioned in Sec.~\ref{sec:measurelf}, if we then add a longitudinal field, the overlap between the dressed ground state and its 1QP sector is non-vanishing, and we require $2(2N+1)$ supplementary circuits. 

For the NLCE calculation on the TFIM chain, we study the model for $N_{\rm max}=5$ on the quantum device, which requires two effective 1QP Hamiltonians on chain segments with 4 and 5 qubits need 230 unique circuits in total. The same $N_{\rm max}$ for the TFIM chain with a longitudinal field requires 303 circuits. Finally, for the ladder TFIM, NLCE+QA with maximal cluster $4 \times 2$ needs 574 circuits.

\subsection{1QP dispersions}

We look at the TFIM on a one-dimensional chain with $h_l=0$ in Eq.~\eqref{eq:ham_tfim} which admits an exact solution via the Jordan-Wigner transformation~\cite{Pfeuty1970}. Assuming $h=1$, the 1QP dispersion $\omega(k)$ at momentum $k$ is given by

\begin{align}
\omega(k) = 2\sqrt{1 + J^2 - 2J\cos(k)}. \label{eq:pfeuty}
\end{align}
The model undergoes a quantum phase transition at critical point $J_c = 1$, separating the quantum paramagnetic phase ($J < 1$) from the symmetry-broken ordered phase ($J > 1$). In this case, the $\mathbb{Z}_2$ symmetry enforces vanishing overlap between the ground-state and the 1QP sector, so the PCAT reduces to the Schrieffer-Wolff transformation. For the 1D TFIM, the rectangular expansion used for NLCE, takes clusters of size $1 \times L_n \leq N_{\text{max}}$. This simplifies NLCE for one-dimensional chain to a subtraction of two consecutive cluster sizes $N_{\mathrm{max}}$ and $N_{\mathrm{max}}-1$, as contributions from smaller clusters cancel. The dispersion $\omega(k)$ in the thermodynamic limit within NLCE taking into account clusters with up to $N_{\mathrm{max}}$ sites, is therefore
\begin{eqnarray}\label{eq:1d_nlce}
    \omega(k) &=& \sum_{\nu,\mu}^{N_{\mathrm{max}}} e^{{\rm i} \, k \,(\nu-\mu)} H^{[1]}_{\mathrm{eff},N_{\mathrm{max}},\nu,\mu}  \\ \nonumber
    &-& \sum_{\nu',\mu'}^{N_{\mathrm{max}}-1}  e^{{\rm i} \, k \, (\nu'-\mu')} H^{[1]}_{\mathrm{eff},N_{\mathrm{max}}-1,\nu'\mu'} ,
\end{eqnarray}
 where $H^{[1]}_{\mathrm{eff}}$ is obtained using Eq.~\eqref{eq:eff_ham_PCAT}.
 
 Since the 1D TFIM admits an exact analytic solution, this case provides a direct benchmark for the hardware results. It is insightful to identify the performance of each part of the algorithm separately. The NLCE itself is an approximation truncated at the largest cluster size, introducing an error because spatial correlations from larger clusters are not considered. The quantum algorithm used will also not perfectly implement the adiabatic connection, whether a VQE or a finite-time sweep is used. The QPU's depolarization noise will further distort the resulting dispersion relation, and the non-linear post-processing due to PCAT amplifies these errors. We investigate these sources of error by adding each layer of error in steps: first, the NLCE is computed with exact Hamiltonian matrices, then with a noiseless simulation of the quantum algorithm (leading to NLCE+QA), then with a noisy simulation, and finally on the QPU.

\begin{figure}[h!]
    \centering
    \includegraphics[width=1\linewidth]{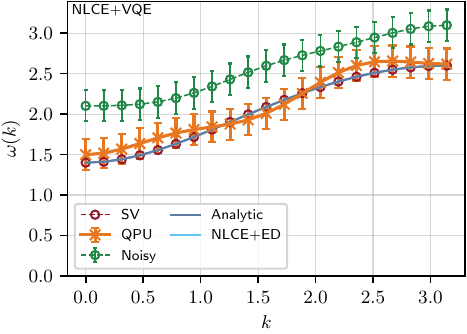}
    \caption{1QP Dispersion for the one-dimensional TFIM chain in the thermodynamic limit for \mbox{$J/h=0.3$} using NLCE+VQE with \mbox{$N_{\mathrm{max}}=5$}. \textbf{NCLE+ED} indicates NLCE performed with exact diagonalization as a cluster solver; it shows the best possible performance of NLCE when using these cluster sizes. \textbf{Analytic} labels the analytic solution of the one-dimensional TFIM model, given in Eq.~\eqref{eq:pfeuty}. \textbf{SV} indicates a noiseless statevector simulation. \textbf{Noisy} indicates a simulation with gate-level depolarization noise with rates set to be close to an AQT device. \textbf{QPU} indicates an experimental result on a 20-qubit AQT trapped-ion device.  The error bars denote one standard deviation of statistical uncertainty.}
    \label{fig:tfim_30j_qpu_results_vqe}
\end{figure}

We now look at the NLCE+VQE setup described above to compute the dispersion in the thermodynamic limit. After the VQE is trained on classical resources, we compute all the required matrix elements on a QPU, and then perform PCAT and NLCE. Fig.~\ref{fig:tfim_30j_qpu_results_vqe} shows the results of noiseless, noisy, and QPU runs for the TFIM chain at $J/h=0.3$. Right away, we notice that the noiseless dispersion overlays the analytic one, indicating that we should not expect error from the VQE being inaccurate. Secondly, the noisy simulation does not overlap with the analytic solution, for the reasons we discussed previously. It uses the reference depolarization rates, reproducing the behaviour in the second column of Fig.~\ref{fig:depol_study}. Interestingly, the QPU dispersion in fact provides a better approximation than the noisy simulation would have suggested. Again, referring to Fig.~\ref{fig:depol_study}, we see that the effective depolarization rate suggested by the QPU dispersion is closer to the third column in that figure,  which may indicate that the reference depolarization rates overestimate the experimental ones. These dispersion curves were obtained with 2000 shots for each matrix element, corresponding to the second row of Fig.~\ref{fig:depol_study}. Hence, we should expect that with more sampling, the QPU should model the dispersion very accurately. 

We can error-mitigate the results by post-selecting the results. Since the preparation circuit for the TFIM without longitudinal field conserves the parity of the excitation number, we can conserve only odd parity outcomes. This would however reduce the shot count further and was therefore not performed. In fact, noisy simulations show that the net effect of loss in statistics due to post-selection actually increases the inaccuracy of the overall result. 

Again using the conserved parity of the excitation number, we can also infer a depolarizing rate $p_O$ from that symmetry, as discussed in Sec.~\ref{sec:error_analysis}. Still, this results in further complications as the overall parity observable is global, so that its rate is different from that of the Hamiltonian terms~\cite{granet_Dilution_2025}. Using more advanced techniques~\cite{delmoral_2026} would require an unacceptable sampling overhead, considering that the rate needs to be computed for each matrix element, meaning we must allocate a significant part of the shot budget to simply compute the depolarization rate. We have therefore not implemented it. 

\begin{figure}
    \centering
    \includegraphics[width=1\linewidth]{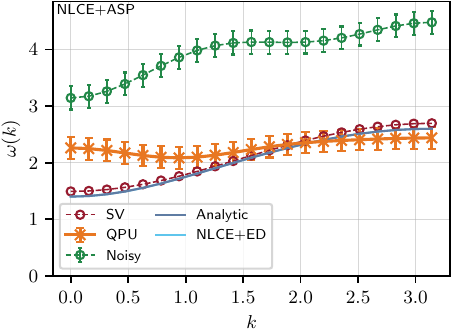}
    \caption{1QP dispersion for the one-dimensional TFIM model in the thermodynamic limit at $J/h=0.3$, obtained with NLCE+ASP. The results take $N_{\mathrm{max}}=5$ in NLCE calculation, and the sweep is done in ten time steps. Labels are the same as in Fig.~\ref{fig:tfim_30j_qpu_results_vqe}. The error bars denote one standard deviation of statistical uncertainty.}
    \label{fig:TFIM_30j_0hl_results_sweep}
\end{figure}

Now, in Fig.~\ref{fig:TFIM_30j_0hl_results_sweep} we show an NLCE+ASP run on the QPU for the same model as discussed before, a pure TFIM chain at $J/h=0.3$ in the thermodynamic limit with $N_{\mathrm{max}}=5$. 
Before we proceed, let us recall that the sweeping time need not be adiabatic with respect to single eigenlevels, but rather at an approximate subspace level, that is between the 1QP sector and the surrounding levels. Therefore, we have heuristically picked, for the results of Fig.~\ref{fig:TFIM_30j_0hl_results_sweep}, 10 time steps, each of which equal to $ 1/(\abs{H_0} + \abs{H})$, with the total sweep time equal to the number of time steps, as described by~\textcite{an_Large_2025}. We can see that this number of steps suffices to reproduce the dispersion relations in the noiseless case. In App.~\ref{app:goodsweep} we show simulations that motivate our choice.

Again, the noisy simulations are overestimating the depolarization rate in the device, and the QPU dispersion curve is closer to the analytic solution than the noisy simulation, indicating an actual depolarization rate closer, again, to the third column of Fig.~\ref{fig:depol_study}. The ASP circuit, however, is 10 times deeper than the VQE circuit, causing a much higher noisy simulation dispersion curve, and a worse QPU result. We note that the QPU runs with higher $J/h$ ratio adhere much closer to the noise model, as discussed in App.~\ref{app:detailedres}.

\begin{figure}
    \centering
    \includegraphics[width=1\linewidth]{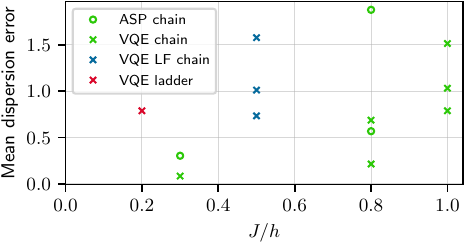}
    \caption{Mean dispersion error, $|\omega_{\text{NLCE+QA}}(k) - \omega_{\text{NLCE+ED}}(k)|$ averaged over $k$, as a function of $J/h$. \textbf{ASP chain}: NLCE+ASP for a TFIM chain with $N_{\mathrm{max}}=5$ at $J/h=0.3, 0.8$. \textbf{VQE chain}: NLCE+VQE for a TFIM chain with $N_{\mathrm{max}}=5$ at $J/h=0.3, 0.8$ and at the critical point at $J/h=1$. \textbf{VQE LF chain}: NLCE+VQE for the TFIM+LF chain with $N_{\mathrm{max}}=5$ at $h_l=0.1$ and $J/h=0.5$. \textbf{VQE ladder}: NLCE+VQE for the TFIM on a ladder geometry at $J/h=0.2$ with $N_{\mathrm{max}}=8$ using clusters of sizes $4 \times 2$ and $3 \times 2$. Multiple markers at the same $J/h$ value indicate different QPU runs that were taken over a period of several weeks.}
    \label{fig:allerrorsqpu}
\end{figure}

Going closer to the quantum-critical point $J/h=1$ for 1D TFIM, adding on a longitudinal field to it, or generally studying more complex models, the error rises as the norms of off-diagonal elements in the matrices grow. Fig.~\ref{fig:allerrorsqpu} displays the mean error of the dispersion curve with respect to the analytic solution (for pure TFIM), and for non-integrable models (TFIM on ladder and TFIM+LF on a chain), to the curve computed using exact diagonalization, as a function of the ratio $J/h$, while also including qualitatively different regimes, which are indicated by different colors. This graph presents a summary of multiple runs over multiple regimes, for which we show detailed presentations in App.~\ref{app:detailedres}.

The pure TFIM chain is shown in green, and we recognize at $J/h=0.3$ the curves shown in Figs.~\ref{fig:tfim_30j_qpu_results_vqe} and~\ref{fig:TFIM_30j_0hl_results_sweep}. The error grows as the value of $J/h$ gets larger and moves closer to the critical point at $J/h=1$. Because of the shallower circuit, the VQE almost always outperforms ASP except in a single case for $J/h=0.8$. It is clear, however, that the mean error over multiple ASP runs should lie much higher than that for VQE runs, provided more data were available to compute a meaningful average. Indeed, we did not average the quantities in Fig.~\ref{fig:allerrorsqpu} because there are too few runs for the statistical mean to be properly representative of the true performance of different algorithms. Still, we note the fairly large difference between the different runs: this underlines the large variance in different realizations of the dispersion, which is due to the non-linear post-processing, as we discussed in Sec.~\ref{sec:error_analysis}, and the fact that the data acquisition phase took place over a period of several weeks.

The orange crosses represent the error for the TFIM with a longitudinal field $h_l=0.1$ at $J/h=0.5$ as discussed in Sec.~\ref{sec:tfim}; this field breaks the $\mathbb{Z}_2$ symmetry, and the PCAT correction becomes fully utilized. Because the off-diagonal matrix elements are larger compared to the diagonal, the inversion is more sensitive, and the overall error larger.

We also consider the TFIM on a two-leg ladder at $J/h = 0.2$, the simplest two-dimensional geometry accessible with our cluster sizes. The ladder has a two-site unit cell, yielding two 1QP bands. Since the ladder is quasi one-dimensional, the rectangular NLCE expansion at $N_{\mathrm{max}}=8$ reduces to two consecutive cluster sizes, here $4 \times 2$ and $3 \times 2$, as in the chain case Eq.~\eqref{eq:1d_nlce}. We present detailed results for this model in App.~\ref{app:ladder}. In Fig.~\ref{fig:allerrorsqpu}, the red crosses correspond to NLCE calculations for the TFIM ladder with the aforementioned cluster sizes. Despite the small value $J/h=0.2$, the error is larger than the one of the TFIM chain for $J/h=0.3$. This can be attributed to the larger qubit count and higher connectivity of the ladder geometry, and the fact that $J/h = 0.2$ is relatively closer to the ladder's critical point $J/h = 0.545$~\cite{ladder_critical} than $J/h = 0.3$ is to the chain's critical point $J/h = 1$.

\section{Conclusions}\label{sec:outlook}

In this work, we explored the dispersion relations of transverse-field Ising models in the polarized phase using the NLCE+QA method, with expectation values computed on a 20-qubit AQT trapped-ion QPU. We introduced a simpler variant of the Hadamard test, the CX-test, which is particularly useful when computing a polynomial amount of matrix elements of a unitary in the computational basis. Using this test in combination with a specifically-trained VQE, we showed dispersion relations with better accuracy than the ASP, but we conjecture that decreasing error rates by a factor $10$ to $100$ should enable accurate computation of the dispersion relations with ASP. Using noisy simulations, we explored the effects of depolarizing noise on non-linear post-processing, to investigate the relationship between depolarization noise, shot noise, and their effect on the NLCE+QA framework.

While the computed dispersion does not always perfectly overlap the analytic dispersion, it can be made to approach it successfully for moderate Ising interactions in the polarized phase, sufficiently far from the quantum-critical point, showing that QPUs can generate data that is reliable enough for non-linear classical post-processing. Importantly, the PCAT construction guarantees cluster additivity of the final effective Hamiltonian on each cluster irrespective of the quantum algorithm used to prepare the input states. This means that different clusters within the NLCE expansion may be solved using different quantum algorithms, for instance VQE for some and ASP for others, while the resulting effective Hamiltonian remains cluster-additive by construction.

The full PCAT correction beyond the Schrieffer-Wolff transformation, which is relevant when the $\mathbb{Z}_2$ symmetry of the TFIM is broken by the longitudinal field, could not be decisively tested at the current level of QPU precision. Errors in individual expectation values are already amplified through the non-linear processing inherent in the Schrieffer-Wolff transformation. As discussed in Sec.~\ref{sec:error_analysis}, the correction $\Delta_{ij} \ll h/\sqrt{M}$ is below the sampling-noise floor at the present shot budget. Demonstrating this correction would thus require more shots and lower hardware error rates. 

As QPUs improve both in terms of noise rate and sampling rate, it may be possible to increase cluster sizes to obtain better approximations of the dispersion, and, eventually, compute the dispersion of models beyond classical means. We have shown in Fig.~\ref{fig:n_circ_needed} that for a $10 \times 10$ square lattice, the NLCE+QA procedure would require around $3\cdot 10^6$ unique circuits, which at current QPU speeds, would require on the order of $10^5$ hours of QPU time, underscoring the need for algorithmic improvements and faster sampling rates. Running multiple circuits in parallel on the quantum device is a straightforward improvement, and the reuse of some expectation values could further bring the number of circuits down. 

Finally, determining how to mitigate error under the assumption of depolarizing noise seems to be a fruitful endeavor, and understanding the low-sampling edge cases when depolarizing noise can almost be assumed, but is not necessarily resolvable. Then, Pauli noise is not homogeneous and analysis becomes non-trivial. \\

\section{Acknowledgements}

The authors gratefully acknowledge the use of the quantum system Marmot by AQT, operated by the Leibniz Supercomputing Centre (LRZ) in Garching, Germany, for providing the computational resources for this work. The authors acknowledge the support by the Munich Quantum Valley, which is supported by the Bavarian state government with funds from the Hightech Agenda Bayern Plus. The authors thank Thomas Monz and Max Hörmann for useful discussions.

\bibliographystyle{apsrev4-2}
\bibliography{references}

@article{Hormann2023a,
  title = {Projective Cluster-Additive Transformation for Quantum Lattice Models},
  author = {H{\"o}rmann, Max and Schmidt, Kai Phillip},
  year = 2023,
  month = sep,
  journal = {SciPost Physics},
  volume = {15},
  number = {3},
  pages = {097},
  issn = {2542-4653},
  doi = {10.21468/SciPostPhys.15.3.097},
  urldate = {2025-12-10},
  langid = {english}
}

@article{Sumeet2024,
  title = {Hybrid Quantum-Classical Algorithm for the Transverse-Field {{Ising}} Model in the Thermodynamic Limit},
  author = {{Sumeet} and H{\"o}rmann, M. and Schmidt, K. P.},
  year = 2024,
  month = oct,
  journal = {Physical Review B},
  volume = {110},
  number = {15},
  pages = {155128},
  publisher = {American Physical Society},
  doi = {10.1103/PhysRevB.110.155128},
  urldate = {2025-12-10}
}

@article{arute_Quantum_2019,
  title = {Quantum Supremacy Using a Programmable Superconducting Processor},
  author = {Arute, Frank and Arya, Kunal and Babbush, Ryan and Bacon, Dave and Bardin, Joseph C. and Barends, Rami and Biswas, Rupak and Boixo, Sergio and Brandao, Fernando G. S. L. and Buell, David A. and Burkett, Brian and Chen, Yu and Chen, Zijun and Chiaro, Ben and Collins, Roberto and Courtney, William and Dunsworth, Andrew and Farhi, Edward and Foxen, Brooks and Fowler, Austin and Gidney, Craig and Giustina, Marissa and Graff, Rob and Guerin, Keith and Habegger, Steve and Harrigan, Matthew P. and Hartmann, Michael J. and Ho, Alan and Hoffmann, Markus and Huang, Trent and Humble, Travis S. and Isakov, Sergei V. and Jeffrey, Evan and Jiang, Zhang and Kafri, Dvir and Kechedzhi, Kostyantyn and Kelly, Julian and Klimov, Paul V. and Knysh, Sergey and Korotkov, Alexander and Kostritsa, Fedor and Landhuis, David and Lindmark, Mike and Lucero, Erik and Lyakh, Dmitry and Mandr{\`a}, Salvatore and McClean, Jarrod R. and McEwen, Matthew and Megrant, Anthony and Mi, Xiao and Michielsen, Kristel and Mohseni, Masoud and Mutus, Josh and Naaman, Ofer and Neeley, Matthew and Neill, Charles and Niu, Murphy Yuezhen and Ostby, Eric and Petukhov, Andre and Platt, John C. and Quintana, Chris and Rieffel, Eleanor G. and Roushan, Pedram and Rubin, Nicholas C. and Sank, Daniel and Satzinger, Kevin J. and Smelyanskiy, Vadim and Sung, Kevin J. and Trevithick, Matthew D. and Vainsencher, Amit and Villalonga, Benjamin and White, Theodore and Yao, Z. Jamie and Yeh, Ping and Zalcman, Adam and Neven, Hartmut and Martinis, John M.},
  year = 2019,
  month = oct,
  journal = {Nature},
  volume = {574},
  number = {7779},
  pages = {505--510},
  publisher = {Nature Publishing Group},
  issn = {1476-4687},
  doi = {10.1038/s41586-019-1666-5},
  urldate = {2024-10-28},
  copyright = {2019 The Author(s), under exclusive licence to Springer Nature Limited},
  langid = {english}
}

@article{friedrich_Precision_2018,
  title = {Precision Matrix Expansion - Efficient Use of Numerical Simulations in Estimating Errors on Cosmological Parameters},
  author = {Friedrich, Oliver and Eifler, Tim},
  year = 2018,
  month = jan,
  journal = {Monthly Notices of the Royal Astronomical Society},
  volume = {473},
  number = {3},
  eprint = {1703.07786},
  primaryclass = {astro-ph},
  pages = {4150--4163},
  issn = {0035-8711, 1365-2966},
  doi = {10.1093/mnras/stx2566},
  urldate = {2025-12-10},
  archiveprefix = {arXiv}
}

@misc{schiffer_Hardwareefficient_2025,
  title = {Hardware-Efficient Quantum Phase Estimation via Local Control},
  author = {Schiffer, Benjamin F. and Wild, Dominik S. and Maskara, Nishad and Lukin, Mikhail D. and Cirac, J. Ignacio},
  year = 2025,
  month = sep,
  number = {arXiv:2506.18765},
  eprint = {2506.18765},
  primaryclass = {quant-ph},
  publisher = {arXiv},
  doi = {10.48550/arXiv.2506.18765},
  urldate = {2025-10-01},
  archiveprefix = {arXiv}
}

@article{pfeuty1970,
  author = {Pfeuty, P.},
  title = {The one-dimensional {Ising} model with a transverse field},
  journal = {Annals of Physics},
  volume = {57},
  number = {1},
  pages = {79--90},
  year = {1970},
  doi = {10.1016/0003-4916(70)90270-8}
}

@misc{sumeet2025,
  author={Sumeet and M. Hörmann and K. P. Schmidt},
  title={Quantum algorithm for one quasi-particle excitations in the thermodynamic limit via cluster-additive block-diagonalization},
  year=2025,
  eprint={2511.06623},
  archivePrefix={arXiv},
  primaryClass={quant-ph},
  url={https://arxiv.org/abs/2511.06623}, 
}

@article{albash_Adiabatic_2018,
  title = {Adiabatic Quantum Computation},
  author = {Albash, Tameem and Lidar, Daniel A.},
  year = 2018,
  month = jan,
  journal = {Reviews of Modern Physics},
  volume = {90},
  number = {1},
  pages = {015002},
  publisher = {American Physical Society},
  doi = {10.1103/RevModPhys.90.015002},
  urldate = {2024-10-21}
}

@article{granet_Dilution_2025,
  title = {Dilution of {{Error}} in {{Digital Hamiltonian Simulation}}},
  author = {Granet, Etienne and Dreyer, Henrik},
  date = {2025-02-24},
  journal = {PRX Quantum},
  volume = {6},
  number = {1},
  pages = {010333},
  publisher = {American Physical Society},
  doi = {10.1103/PRXQuantum.6.010333},
  url = {https://link.aps.org/doi/10.1103/PRXQuantum.6.010333},
  year=2025,
  month=feb
}

@misc{delmoral_2026,
  title = {Noise Mitigation of Quantum Observables via Learning from {{Hamiltonian}} Symmetry Decays},
  author = {del Moral, Javier Oliva and Larrarte, Olatz Sanz and Fraxanet, Joana and Mishagli, Dmytro and Martinez, Josu Etxezarreta},
  date = {2026},
  doi = {10.48550/ARXIV.2603.13060},
  url = {https://arxiv.org/abs/2603.13060},
  urldate = {2026-05-18},
  pubstate = {prepublished},
  version = {1},
  year = 2026
}

@misc{qiskit,
  title = {Quantum Computing with {{Qiskit}}},
  author = {{Javadi-Abhari}, Ali and Treinish, Matthew and Krsulich, Kevin and Wood, Christopher J. and Lishman, Jake and Gacon, Julien and Martiel, Simon and Nation, Paul D. and Bishop, Lev S. and Cross, Andrew W. and Johnson, Blake R. and Gambetta, Jay M.},
  year = 2024,
  month = jun,
  number = {arXiv:2405.08810},
  eprint = {2405.08810},
  primaryclass = {quant-ph},
  publisher = {arXiv},
  doi = {10.48550/arXiv.2405.08810},
  urldate = {2026-01-21},
  archiveprefix = {arXiv}
}

@article{peruzzo_variational_2014,
  title = {A Variational Eigenvalue Solver on a Photonic Quantum Processor},
  author = {Peruzzo, Alberto and McClean, Jarrod and Shadbolt, Peter and Yung, Man-Hong and Zhou, Xiao-Qi and Love, Peter J. and {Aspuru-Guzik}, Al{\'a}n and O'Brien, Jeremy L.},
  year = 2014,
  month = jul,
  journal = {Nature Communications},
  volume = {5},
  number = {1},
  pages = {4213},
  publisher = {Nature Publishing Group},
  issn = {2041-1723},
  doi = {10.1038/ncomms5213},
  urldate = {2022-02-16},
  copyright = {2014 The Author(s)},
  langid = {english}
}

@misc{kanno_QuantumSelected_2023,
  title = {Quantum-{{Selected Configuration Interaction}}: Classical Diagonalization of {{Hamiltonians}} in Subspaces Selected by Quantum Computers},
  shorttitle = {Quantum-{{Selected Configuration Interaction}}},
  author = {Kanno, Keita and Kohda, Masaya and Imai, Ryosuke and Koh, Sho and Mitarai, Kosuke and Mizukami, Wataru and Nakagawa, Yuya O.},
  year = 2023,
  month = feb,
  number = {arXiv:2302.11320},
  eprint = {2302.11320},
  primaryclass = {quant-ph},
  publisher = {arXiv},
  doi = {10.48550/arXiv.2302.11320},
  urldate = {2025-01-31},
  archiveprefix = {arXiv}
}

@book{nielsen_Quantum_2010,
  title = {Quantum {{Computation}} and {{Quantum Information}}: 10th {{Anniversary Edition}}},
  shorttitle = {Quantum {{Computation}} and {{Quantum Information}}},
  author = {Nielsen, Michael A. and Chuang, Isaac L.},
  year = 2010,
  month = dec,
  publisher = {Cambridge University Press},
  isbn = {978-1-139-49548-6},
  langid = {english}
}

@misc{larocca_Review_2024,
  title = {A {{Review}} of {{Barren Plateaus}} in {{Variational Quantum Computing}}},
  author = {Larocca, Martin and Thanasilp, Supanut and Wang, Samson and Sharma, Kunal and Biamonte, Jacob and Coles, Patrick J. and Cincio, Lukasz and McClean, Jarrod R. and Holmes, Zo{\"e} and Cerezo, M.},
  year = 2024,
  month = may,
  number = {arXiv:2405.00781},
  eprint = {2405.00781},
  primaryclass = {quant-ph, stat},
  publisher = {arXiv},
  doi = {10.48550/arXiv.2405.00781},
  urldate = {2024-08-20},
  archiveprefix = {arXiv}
}

@misc{nutzel_Ground_2025,
  title = {Ground {{State Energy}} via {{Adiabatic Evolution}} and {{Phase Measurement}} for a {{Molecular Hamiltonian}} on an {{Ion-Trap Quantum Computer}}},
  author = {N{\"u}tzel, Ludwig and Hartmann, Michael J. and Dreyer, Henrik and Granet, Etienne},
  year = 2025,
  month = dec,
  number = {arXiv:2512.14415},
  eprint = {2512.14415},
  primaryclass = {quant-ph},
  publisher = {arXiv},
  doi = {10.48550/arXiv.2512.14415},
  urldate = {2026-01-30},
  archiveprefix = {arXiv}
}

@misc{an_Large_2025,
  title = {Large Time-Step Discretisation of Adiabatic Quantum Dynamics},
  author = {An, Dong and Costa, Pedro C. S. and Berry, Dominic W.},
  year = 2025,
  month = aug,
  number = {arXiv:2509.00171},
  eprint = {2509.00171},
  primaryclass = {quant-ph},
  publisher = {arXiv},
  doi = {10.48550/arXiv.2509.00171},
  urldate = {2025-09-03},
  archiveprefix = {arXiv}
}

@book{GUM_2008,
  title = {Evaluation of Measurement Data — {{Guide}} to the Expression of Uncertainty in Measurement},
  date = {2008},
  doi = {10.59161/JCGM100-2008E},
  url = {https://www.bipm.org/doi/10.59161/JCGM100-2008E},
  urldate = {2026-04-16},
  year = 2008,
  author = {Joint Committee for Guides in Metrology},
  publisher = {International Organization for Standardization}
  
}

@misc{zhang_Modern_2020,
  title = {Modern {{Monte Carlo Methods}} for {{Efficient Uncertainty Quantification}} and {{Propagation}}: {{A Survey}}},
  shorttitle = {Modern {{Monte Carlo Methods}} for {{Efficient Uncertainty Quantification}} and {{Propagation}}},
  author = {Zhang, Jiaxin},
  year = 2020,
  month = nov,
  number = {arXiv:2011.00680},
  eprint = {2011.00680},
  primaryclass = {stat},
  publisher = {arXiv},
  doi = {10.48550/arXiv.2011.00680},
  urldate = {2026-01-30},
  archiveprefix = {arXiv}
}

@inproceedings{wan_unscented_2000,
  title = {The Unscented {{Kalman}} Filter for Nonlinear Estimation},
  booktitle = {Proceedings of the {{IEEE}} 2000 {{Adaptive Systems}} for {{Signal Processing}}, {{Communications}}, and {{Control Symposium}} ({{Cat}}. {{No}}.{{00EX373}})},
  author = {Wan, E.A. and Van Der Merwe, R.},
  year = 2000,
  month = oct,
  pages = {153--158},
  doi = {10.1109/ASSPCC.2000.882463},
  urldate = {2025-12-02}
}

@misc{romano_Fundamental_2025,
  title = {Fundamental {{Topics}} in {{Continuum Mechanics}}: {{Grand Ideas}}, {{Errors}} \& {{Horrors}}},
  shorttitle = {Fundamental {{Topics}} in {{Continuum Mechanics}}},
  author = {Romano, Giovanni and Barretta, Raffaele},
  year = 2025,
  month = nov,
  number = {arXiv:2511.08129},
  eprint = {2511.08129},
  primaryclass = {physics},
  publisher = {arXiv},
  doi = {10.48550/arXiv.2511.08129},
  urldate = {2026-02-02},
  archiveprefix = {arXiv}
}

@article{wang_New_2024,
  title = {New {{Quantum Algorithms}} for {{Computing Quantum Entropies}} and {{Distances}}},
  author = {Wang, Qisheng and Guan, Ji and Liu, Junyi and Zhang, Zhicheng and Ying, Mingsheng},
  year = 2024,
  month = aug,
  journal = {IEEE Transactions on Information Theory},
  volume = {70},
  number = {8},
  eprint = {2203.13522},
  primaryclass = {quant-ph},
  pages = {5653--5680},
  issn = {0018-9448, 1557-9654},
  doi = {10.1109/TIT.2024.3399014},
  urldate = {2026-02-02},
  archiveprefix = {arXiv}
}

@article{demko_1984,
  title = {Decay Rates for Inverses of Band Matrices},
  author = {Demko, Stephen and Moss, William F. and Smith, Philip W.},
  date = {1984},
  year = 1984,
  journal = {Mathematics of Computation},
  shortjournal = {Math. Comp.},
  volume = {43},
  number = {168},
  pages = {491--499},
  issn = {0025-5718, 1088-6842},
  doi = {10.1090/S0025-5718-1984-0758197-9},
  url = {https://www.ams.org/mcom/1984-43-168/S0025-5718-1984-0758197-9/},
  urldate = {2026-04-05},
  langid = {english}
}

@misc{eisert_Mind_2025,
  title = {Mind the Gaps: {{The}} Fraught Road to Quantum Advantage},
  shorttitle = {Mind the Gaps},
  author = {Eisert, Jens and Preskill, John},
  year = 2025,
  month = nov,
  number = {arXiv:2510.19928},
  eprint = {2510.19928},
  primaryclass = {quant-ph},
  publisher = {arXiv},
  doi = {10.48550/arXiv.2510.19928},
  urldate = {2026-02-03},
  archiveprefix = {arXiv}
}

@article{urbanek_Mitigating_2021,
  title = {Mitigating Depolarizing Noise on Quantum Computers with Noise-Estimation Circuits},
  author = {Urbanek, Miroslav and Nachman, Benjamin and Pascuzzi, Vincent R. and He, Andre and Bauer, Christian W. and de Jong, Wibe A.},
  date = {2021-12-27},
  year=2021,
  journal = {Physical Review Letters},
  shortjournal = {Phys. Rev. Lett.},
  volume = {127},
  number = {27},
  eprint = {2103.08591},
  eprinttype = {arXiv},
  eprintclass = {quant-ph},
  pages = {270502},
  issn = {0031-9007, 1079-7114},
  doi = {10.1103/PhysRevLett.127.270502},
  url = {http://arxiv.org/abs/2103.08591},
  urldate = {2026-03-11}
}

@article{cai_Quantum_2023,
  title = {Quantum {{Error Mitigation}}},
  author = {Cai, Zhenyu and Babbush, Ryan and Benjamin, Simon C. and Endo, Suguru and Huggins, William J. and Li, Ying and McClean, Jarrod R. and O'Brien, Thomas E.},
  date = {2023-12-13},
  year = 2023,
  journal = {Reviews of Modern Physics},
  shortjournal = {Rev. Mod. Phys.},
  volume = {95},
  number = {4},
  eprint = {2210.00921},
  eprinttype = {arXiv},
  eprintclass = {quant-ph},
  pages = {045005},
  issn = {0034-6861, 1539-0756},
  doi = {10.1103/RevModPhys.95.045005},
  url = {http://arxiv.org/abs/2210.00921},
  urldate = {2026-04-07}
  
}

@article{mele_Noiseinduced_2026,
  title = {Noise-Induced Shallow Circuits and the Absence of Barren Plateaus},
  author = {Mele, Antonio Anna and Angrisani, Armando and Ghosh, Soumik and Khatri, Sumeet and Eisert, Jens and Stilck França, Daniel and Quek, Yihui},
  date = {2026-04-02},
  journal = {Nature Physics},
  shortjournal = {Nat. Phys.},
  pages = {1--6},
  publisher = {Nature Publishing Group},
  issn = {1745-2481},
  doi = {10.1038/s41567-026-03245-z},
  url = {https://www.nature.com/articles/s41567-026-03245-z},
  urldate = {2026-04-09},
  year=2026,
  langid = {english}
}

@misc{lischke_What_2019,
  title = {What {{Is}} the {{Fractional Laplacian}}?},
  author = {Lischke, Anna and Pang, Guofei and Gulian, Mamikon and Song, Fangying and Glusa, Christian and Zheng, Xiaoning and Mao, Zhiping and Cai, Wei and Meerschaert, Mark M. and Ainsworth, Mark and Karniadakis, George Em},
  year = 2019,
  month = nov,
  number = {arXiv:1801.09767},
  eprint = {1801.09767},
  primaryclass = {math},
  publisher = {arXiv},
  doi = {10.48550/arXiv.1801.09767},
  urldate = {2026-02-03},
  archiveprefix = {arXiv}
}

@article{lefebvre_Propagation_2000,
  title = {Propagation of {{Errors}} for {{Matrix Inversion}}},
  author = {Lefebvre, M. and Keeler, R. K. and Sobie, R. and White, J.},
  year = 2000,
  month = sep,
  journal = {Nuclear Instruments and Methods in Physics Research Section A: Accelerators, Spectrometers, Detectors and Associated Equipment},
  volume = {451},
  number = {2},
  eprint = {hep-ex/9909031},
  pages = {520--528},
  issn = {01689002},
  doi = {10.1016/S0168-9002(00)00323-5},
  urldate = {2025-12-02},
  archiveprefix = {arXiv}
}

@book{lakes_Viscoelastic_2009,
  title = {Viscoelastic {{Materials}}},
  author = {Lakes, Roderic},
  year = 2009,
  publisher = {Cambridge University Press},
  address = {Cambridge},
  doi = {10.1017/CBO9780511626722},
  urldate = {2026-02-02},
  isbn = {978-0-521-88568-3}
}

@misc{_MARMOT_,
  title = {{{MARMOT}} - {{The}} First Commercial 19-Inch Rack-Mounted Quantum Computer},
  journal = {AQT - Alpine Quantum Technologies},
  urldate = {2026-02-11},
  howpublished = {https://www.aqt.eu/products/marmot/},
  langid = {american},
  year = 2026
}

@article{huang_Predicting_2020,
  title = {Predicting Many Properties of a Quantum System from Very Few Measurements},
  author = {Huang, Hsin-Yuan and Kueng, Richard and Preskill, John},
  year = 2020,
  month = oct,
  journal = {Nature Physics},
  volume = {16},
  number = {10},
  pages = {1050--1057},
  publisher = {Nature Publishing Group},
  issn = {1745-2481},
  doi = {10.1038/s41567-020-0932-7},
  urldate = {2024-04-22},
  copyright = {2020 The Author(s), under exclusive licence to Springer Nature Limited},
  langid = {english}
}

@article{nlce_Rigol,
  title = {Numerical Linked-Cluster Approach to Quantum Lattice Models},
  author = {Rigol, Marcos and Bryant, Tyler and Singh, Rajiv R. P.},
  journal = {Phys. Rev. Lett.},
  volume = {97},
  issue = {18},
  pages = {187202},
  numpages = {4},
  year = {2006},
  month = {Nov},
  publisher = {American Physical Society},
  doi = {10.1103/PhysRevLett.97.187202},
  url = {https://link.aps.org/doi/10.1103/PhysRevLett.97.187202}
}

@article{TANG2013557,
title = {A short introduction to numerical linked-cluster expansions},
journal = {Computer Physics Communications},
volume = {184},
number = {3},
pages = {557-564},
year = {2013},
issn = {0010-4655},
doi = {https://doi.org/10.1016/j.cpc.2012.10.008},
url = {https://www.sciencedirect.com/science/article/pii/S0010465512003414},
author = {Baoming Tang and Ehsan Khatami and Marcos Rigol}
}

@Article{Peruzzo2014,
author={Peruzzo, Alberto
and McClean, Jarrod
and Shadbolt, Peter
and Yung, Man-Hong
and Zhou, Xiao-Qi
and Love, Peter J.
and Aspuru-Guzik, Al{\'a}n
and O'Brien, Jeremy L.},
title={A variational eigenvalue solver on a photonic quantum processor},
journal={Nature Communications},
year={2014},
month={Jul},
day={23},
volume={5},
number={1},
pages={4213},
issn={2041-1723},
doi={10.1038/ncomms5213},
url={https://doi.org/10.1038/ncomms5213}
}

@article{wecker_progress_2015,
    title = {Progress towards practical quantum variational algorithms},
    volume = {92},
    issn = {10941622},
    doi = {10.1103/PhysRevA.92.042303},
    number = {4},
    journal = {Physical Review A - Atomic, Molecular, and Optical Physics},
    author = {Wecker, Dave and Hastings, Matthew B. and Troyer, Matthias},
    year = {2015},
    pages = {1--10},
}

@article{Bravyi2011,
  author = {Bravyi, S. and DiVincenzo, D. P. and Loss, D.},
  title = {{Schrieffer-Wolff} transformation for quantum many-body systems},
  journal = {Annals of Physics},
  volume = {326},
  number = {10},
  pages = {2793--2826},
  year = {2011},
  doi = {10.1016/j.aop.2011.06.004}
}

@article{suzuki_Decomposition_1985,
  title = {Decomposition Formulas of Exponential Operators and {{Lie}} Exponentials with Some Applications to Quantum Mechanics and Statistical Physics},
  author = {Suzuki, Masuo},
  year = 1985,
  month = apr,
  journal = {Journal of Mathematical Physics},
  volume = {26},
  number = {4},
  pages = {601--612},
  issn = {0022-2488},
  doi = {10.1063/1.526596},
  urldate = {2024-03-09}
}

@article{ladder_critical,
  title = {Coexistence of spontaneous dimerization and magnetic order in a transverse-field Ising ladder with four-spin interactions},
  author = {Xavier, J. C. and Pereira, R. G. and Nunes, M. E. S. and Plascak, J. A.},
  journal = {Phys. Rev. B},
  volume = {105},
  issue = {2},
  pages = {024430},
  numpages = {9},
  year = {2022},
  month = {Jan},
  publisher = {American Physical Society},
  doi = {10.1103/PhysRevB.105.024430},
  url = {https://link.aps.org/doi/10.1103/PhysRevB.105.024430}
}

@article{ipeps,
  title = {Classical Simulation of Infinite-Size Quantum Lattice Systems in Two Spatial Dimensions},
  author = {Jordan, J. and Or\'us, R. and Vidal, G. and Verstraete, F. and Cirac, J. I.},
  journal = {Phys. Rev. Lett.},
  volume = {101},
  issue = {25},
  pages = {250602},
  numpages = {4},
  year = {2008},
  month = {Dec},
  publisher = {American Physical Society},
  doi = {10.1103/PhysRevLett.101.250602},
  url = {https://link.aps.org/doi/10.1103/PhysRevLett.101.250602}
}

@article{cluster-dmft,
  title = {Quantum cluster theories},
  author = {Maier, Thomas and Jarrell, Mark and Pruschke, Thomas and Hettler, Matthias H.},
  journal = {Rev. Mod. Phys.},
  volume = {77},
  issue = {3},
  pages = {1027--1080},
  numpages = {0},
  year = {2005},
  month = {Oct},
  publisher = {American Physical Society},
  doi = {10.1103/RevModPhys.77.1027},
  url = {https://link.aps.org/doi/10.1103/RevModPhys.77.1027}
}

@misc{wolf2025,
      title={Variational Time Evolution Compression for Solving Impurity Models on Quantum Hardware}, 
      author={Stefan Wolf and Martin Eckstein and Michael J. Hartmann},
      year={2025},
      eprint={2508.10526},
      archivePrefix={arXiv},
      primaryClass={quant-ph},
      url={https://arxiv.org/abs/2508.10526}, 
}

@article{Carleo17,
author = {Giuseppe Carleo  and Matthias Troyer },
title = {Solving the quantum many-body problem with artificial neural networks},
journal = {Science},
volume = {355},
number = {6325},
pages = {602-606},
year = {2017},
doi = {10.1126/science.aag2302},
URL = {https://www.science.org/doi/abs/10.1126/science.aag2302},
eprint = {https://www.science.org/doi/pdf/10.1126/science.aag2302}}

@article{Hartmann19,
  title = {Neural-Network Approach to Dissipative Quantum Many-Body Dynamics},
  author = {Hartmann, Michael J. and Carleo, Giuseppe},
  journal = {Phys. Rev. Lett.},
  volume = {122},
  issue = {25},
  pages = {250502},
  numpages = {6},
  year = {2019},
  month = {Jun},
  publisher = {American Physical Society},
  doi = {10.1103/PhysRevLett.122.250502},
  url = {https://link.aps.org/doi/10.1103/PhysRevLett.122.250502}
}

@misc{stockinger2026,
      title={Sample-Based Quantum Diagonalization with Amplitude Amplification}, 
      author={Nina Stockinger and Ludwig Nützel and Michael J. Hartmann},
      year={2026},
      eprint={2605.02565},
      archivePrefix={arXiv},
      primaryClass={quant-ph},
      url={https://arxiv.org/abs/2605.02565}, 
}

\clearpage

\appendix

\begin{widetext}
    
\section{Detailed breakdown of the results}\label{app:detailedres}

In this section, we present a more detailed view of the results discussed previously, and whose mean errors are shown in Fig.~\ref{fig:allerrorsqpu}. We present all dispersion curves, as well as detailed explanations of the different regimes. We show in Tab.~\ref{tab:tfimcases} a summary of all the QPU runs completed during the data acquisition phase. 

 \begin{table}[h!]
     \centering
     \begin{tabular}[t]{c|cccc}\hline\hline
     Model name & Algorithm & Parameters & Clusters & Number of runs \\\hline
     1D chain & VQE & $J/h=0.3$, $h_l=0$ & $5, 4$ qubits & 1 \\
     1D chain & ASP & $J/h=0.3$, $h_l=0$ & $5, 4$ qubits & 1 \\
      1D chain & VQE & $J/h=0.8$, $h_l=0$ & $5, 4$ qubits & 2 \\
     1D chain & ASP & $J/h=0.8$, $h_l=0$ & $5, 4$ qubits & 2 \\
     1D chain & VQE & $J/h=1$, $h_l=0$ & $5, 4$ qubits & 3 \\
     1D with LF & VQE & $J/h=0.5$, $h_l=0.1$ & $5, 4$  qubits & 3 \\
     2D ladder & VQE & $J/h=0.2$, $h_l=0$ & $4 \times 2, 3 \times 2$ qubits & 1 \\\hline
 \end{tabular}
     \caption{Studied TFIM models and parameter regimes for which at least one QPU run was completed. }
     \label{tab:tfimcases}
 \end{table}

 All the resulting dispersions from each QPU run are displayed in the figures below, in varying shades from orange to brown and compared to analytic results of results of NLCE from statevector simulations.

\end{widetext}

\subsection{TFIM chain closer to the quantum-critical point}

\begin{figure}[h!]
    \centering
        \includegraphics[width=.95\linewidth]{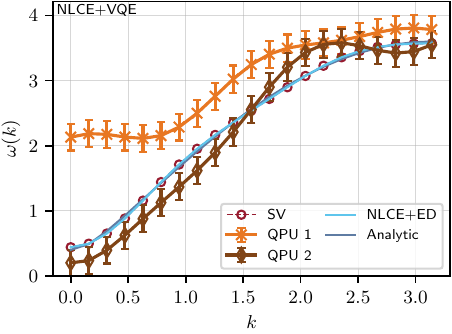}
    \caption{1QP dispersion obtained using NLCE+VQE for the one-dimensional TFIM chain in the thermodynamic limit at \mbox{$J/h=0.8$} using  \mbox{$N_{\mathrm{max}}=5$} for the NLCE expansion. Labels are the same as in Fig.~\ref{fig:tfim_30j_qpu_results_vqe}. The error bars denote one standard deviation of statistical uncertainty, and the orange and brown lines (QPU 1 and QPU 2) show different QPU runs taken several weeks apart.}
    \label{fig:tfim_80j_qpu_results}
\end{figure}

Fig.~\ref{fig:tfim_80j_qpu_results} shows the NLCE+VQE results from two QPU runs together with statevector simulations and analytical results for $J/h=0.8$ in the gapped, disordered phase. In this figure, the difference between the analytic dispersion and the NLCE+ED  is more apparent than in Fig.~\ref{fig:tfim_30j_qpu_results_vqe}; the low amplitudes oscillations around the true dispersion are due to the small size of the clusters. While the first QPU run produces a very close result, the second shows the bias and squeezing expected from the explanation in Sec.~\ref{sec:error_analysis}.

\begin{figure}[h!]
    \centering
    \includegraphics[width=.95\linewidth]{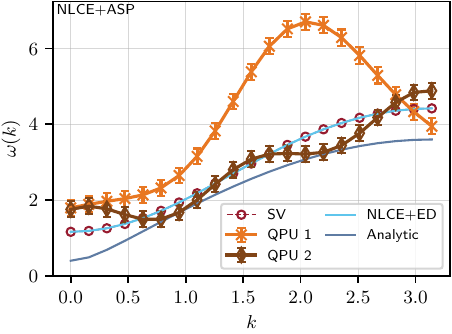}
    \caption{1QP dispersion for the one-dimensional TFIM model in the thermodynamic limit at $J/h=0.8$, obtained with NLCE+ASP. The results take $N_{\mathrm{max}}=5$ in NLCE calculation. Labels are as in Fig.~\ref{fig:tfim_30j_qpu_results_vqe}. The error bars denote one standard deviation of statistical uncertainty, and the orange and brown lines (QPU 1 and QPU 2) show different QPU runs taken several weeks apart.}
    \label{fig:TFIM_80j_0hl_results_sweep}
\end{figure}

We also show the $J/h=0.8$ case using NLCE+ASP in Fig.~\ref{fig:TFIM_80j_0hl_results_sweep}. Again, we use 10 steps, but it is clear that, in this case, it is too low to prevent leaking across the sectors. Still, we have kept this number as a trade-off between accuracy and depth, since a longer sweep would certainly produce a better statevector simulation but a worse QPU result. We note, however, the surprising result in the second run: the QPU almost overlaps the statevector simulation. This can again be explained by the fact that data acquisition was performed over several weeks during which destabilizing phenomena such as stray magnetic fields, miscalibrations, and qubit laser issues may happen, causing significant variation across the different runs.

Regardless, the QPU results show that it is not currently feasible to use ASP in this regime, even with a limited number of steps. However, in experiments where efficient error mitigation is possible, one can successfully perform ASP~\cite{nutzel_Ground_2025}. 

We can extrapolate the error rate required for the QPU by artificially decreasing the error rate in the noisy simulator, to satisfactorily compute the NLCE+ASP pipeline without a particular error-mitigation scheme. We find that a decrease between one and two orders of magnitude in the gate depolarization rates is needed. We investigate this in App.~\ref{app:goodsweep}. Such an improvement is achievable, and is predicted on multiple public developer roadmaps, and, as we noted, the device sometimes produces results which suggest such depolarization rates, though not consistently.

\subsection{TFIM chain at the quantum-critical point} 

\begin{figure}[h!]
    \centering
    \includegraphics[width=.95\linewidth]{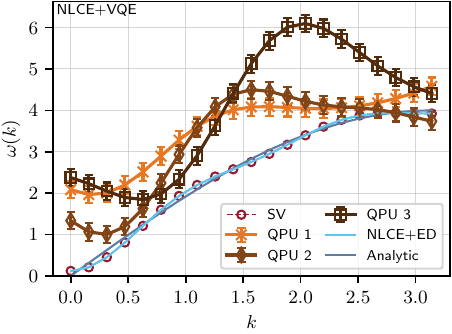}
    \caption{1QP dispersion of the one-dimensional TFIM in the thermodynamic limit at the critical point $J/h=1$ using NLCE+VQE, using $N_{\mathrm{max}}=5$ for the NLCE expansion. Labels are the same as in Fig.~\ref{fig:tfim_30j_qpu_results_vqe}. The error bars denote one standard deviation of statistical uncertainty, and the orange and brown lines (QPU 1, QPU 2, and QPU 3) show different QPU runs taken several weeks apart.}
    \label{fig:tfim_100j_qpu_results}
\end{figure}

Next, we show in Fig.~\ref{fig:tfim_100j_qpu_results} the dispersion curve for \mbox{$J/h=1$} at the quantum-critical point. We observe that the QPU result is even further from the analytical solution. While noise is still the main bottleneck, we see that the noiseless statevector VQE in this case is somewhat further to the analytical solution than that of Fig.~\ref{fig:tfim_30j_qpu_results_vqe}, so the difficulty of that regime makes it harder to compute, mainly because the inverse is more sensitive than the more diagonal $J/h=0.3$ case.

\subsection{TFIM chain with longitudinal field}\label{sec:res_tfim+lf}

The TFIM with longitudinal field (TFIM+LF), whose Hamiltonian is given in Eq.~\eqref{eq:ham_tfim}, is the first benchmark requiring the full PCAT correction. The longitudinal field breaks the $\mathbb{Z}_2$ parity symmetry, coupling the ground state to the 1QP sector and producing non-vanishing overlaps $\langle \Phi^{[0]} | \Psi^{[1]}_i \rangle \neq 0$. As discussed in Sec.~\ref{subsec:PCAT}, this necessitates the PCAT correction to ensure cluster additivity of the NLCE. However, this correction is much smaller for the TFIM+LF in the investigated coupling regime than our sampling budget can observe. Furthermore, as discussed in Sec.~\ref{sec:measurelf}, adding the full PCAT correction requires $2N+1$ more values which must be computed using the CX-test. Therefore, we ignore it and use $\ket{\Psi^{[1]}}$ instead of $\ket{\tilde{\Psi}^{[1]}}$ when computing the overlap matrix $\tilde{O}_{ij}$. 

The absence of parity symmetry also makes the optimization of the variational cost function in Eq.~\eqref{eq:cf_variance} more challenging, as the block-diagonalization can no longer exploit the symmetry constraints.

\begin{figure}[h!]
    \centering
    \includegraphics[width=.95\linewidth]{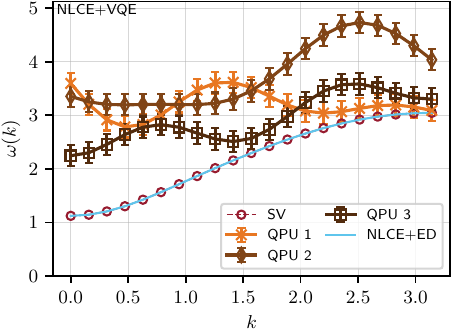}
    \caption{1QP dispersion relation for the one-dimensional TFIM+LF at $J/h=0.5$ and $h_l = 0.1$, in the thermodynamic limit computed with NLCE+VQE, using $N_{\mathrm{max}}=5$. Labels are the same as in Fig.~\ref{fig:tfim_30j_qpu_results_vqe}. The error bars denote one standard deviation of statistical uncertainty, and the orange and brown lines (QPU 1, QPU 2, and QPU 3) show different QPU runs taken several weeks apart.}
    \label{fig:tfim_50j_10hl_qpu_results}
\end{figure}

The results for $J/h = 0.5$, $h_l = 0.1$ are shown in Fig.~\ref{fig:tfim_50j_10hl_qpu_results}. Since the longitudinal field renders the model non-integrable, no analytical solution is available and the NLCE+ED curve with $N_{\rm max}=5$ serves as the reference. The other curves show the results obtained with NLCE+VQE, either using a noiseless simulation or on the QPU.

\begin{figure*}
    \centering
    \includegraphics[width=.75\linewidth]{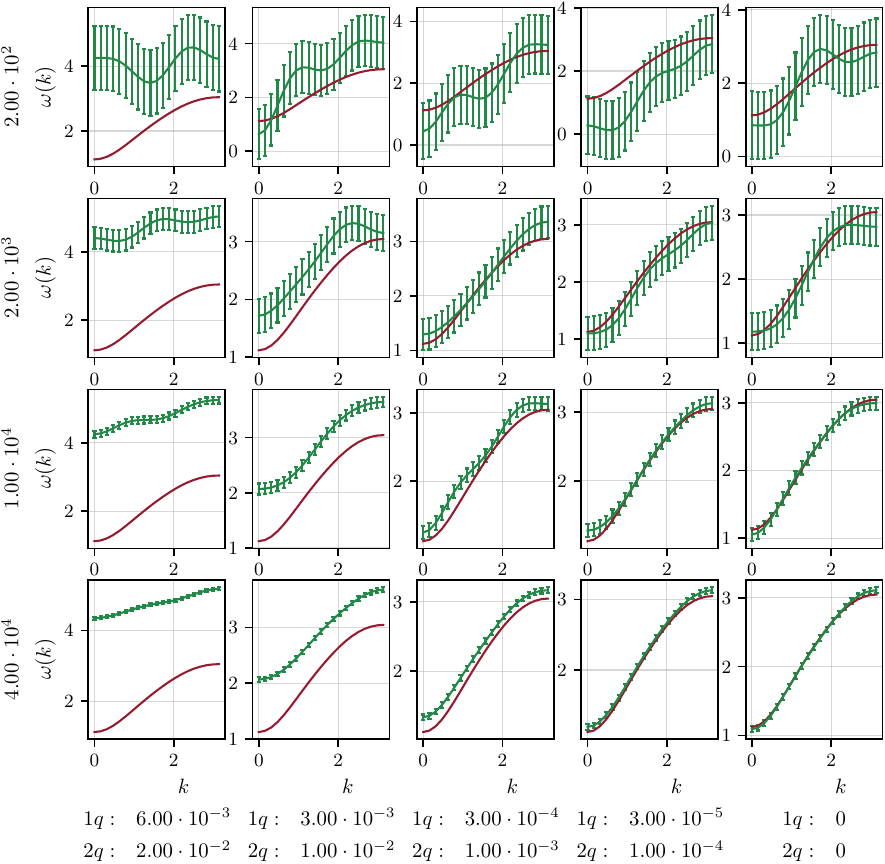}
    \caption{Noisy simulations for multiple shot counts and gate-level depolarization rate, as in Fig.~\ref{fig:depol_study}, in this case for the TFIM+LF chain with $J/h=0.5$ and $h_l=0.1$ obtained by NLCE+QA with $N_{\rm max} = 5$.}
    \label{fig:depol_study+lf}
\end{figure*}

For better context, we show in Fig.~\ref{fig:depol_study+lf} the longitudinal field counterpart to Fig.~\ref{fig:depol_study}, which we recall was done for a TFIM chain with $J/h=0.8$. In this case, we show what happens when we add a longitudinal field. The overall standard deviation is higher, because the shot count per matrix element is fixed, but there is one more term in the Hamiltonian as compared to that of a pure TFIM. This increases the final uncertainty due to the nature of uncertainty propagation. In this regime, the quantum device struggles the most, as is best seen by comparing the second to last column, bottom row panel of Fig.~\ref{fig:depol_study} to this one. It is clear that even with a decrease of depolarization noise by a factor of a hundred, the bias derived in Sec.~\ref{sec:error_analysis} remains visible. Yet, Fig.~\ref{fig:tfim_50j_10hl_qpu_results} indicates that the VQE should match the dispersion closely. We can conclude that the longitudinal field requires higher accuracy than other regimes.

\subsection{TFIM on a ladder}\label{app:ladder}
Finally, we turn to the TFIM on a two-leg ladder, the simplest two-dimensional geometry accessible with our cluster sizes. The Hamiltonian for this model is given by Eq.~\eqref{eq:ham_tfim} for $h_l=0$, with the nearest-neighbor couplings along both rungs and legs of the ladder. Unlike the one-dimensional TFIM, the ladder does not have an exact analytical solution. The model undergoes a quantum phase transition between the paramagnetic and the ferromagnetic phase at $J/h\approx 0.545$~\cite{ladder_critical}. Since the ladder is quasi one-dimensional, the aforementioned quantum phase transition belongs to the 2D Ising universality class. 

\begin{figure}
    \centering
    \includegraphics[width=.95\linewidth]{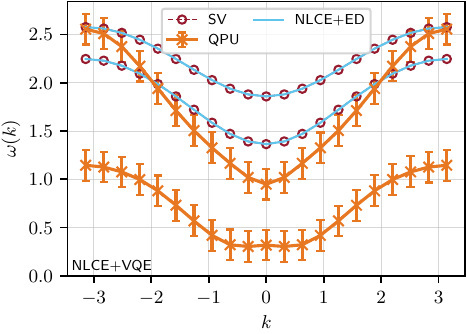}
    \caption{1QP dispersion of the quasi one-dimensional TFIM ladder in the thermodynamic limit at \mbox{$J/h=0.2$} using NLCE+VQE with \mbox{$N_{\mathrm{max}}=8$}. Labels are the same as in Fig.~\ref{fig:tfim_30j_qpu_results_vqe}. The error bars denote one standard deviation of statistical uncertainty.}
    \label{fig:tfim_ladder}
\end{figure}

The two-site unit cell makes this the only benchmark where the 1QP dispersion consists of multiple bands. The Fourier transform of the effective Hamiltonian yields a $2 \times 2$ matrix at each momentum $k$, whose eigenvalues give the bonding and anti-bonding dispersions. At the quantum phase transition, the gap closes at $k = 0$ in the lower (bonding) band, while the upper (anti-bonding) band remains gapped throughout. The NLCE for the ladder with $N_{\mathrm{max}}=8$ requires $4 \times 2$ and $3 \times 2$ clusters. We work at $J/h = 0.2$, in the paramagnetic phase, where the excitation gap is large and the NLCE converges rapidly with cluster size, allowing us to isolate hardware-induced errors from NLCE truncation effects. The results for this geometry are shown in Fig.~\ref{fig:tfim_ladder}. 

\subsection{Matrix elements}
\begin{figure*}
    \centering
    \includegraphics[width=.95\linewidth]{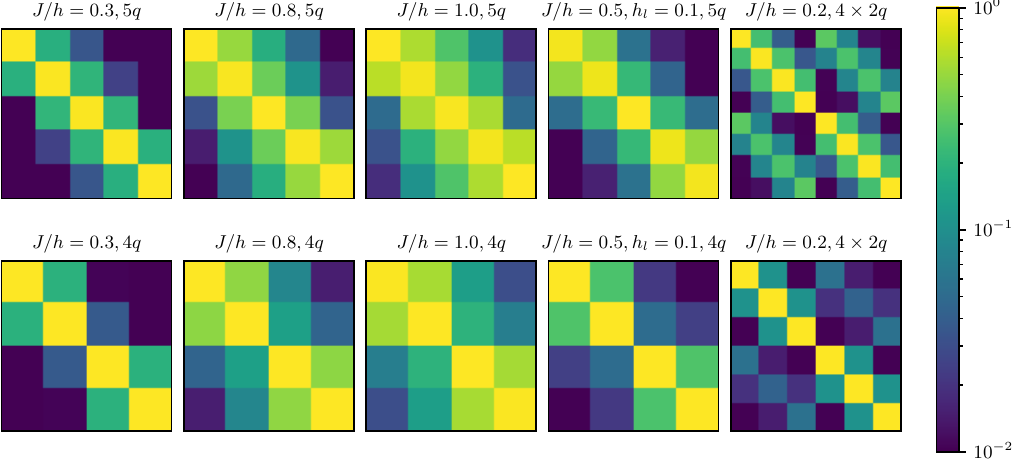}
    \caption{Normalized moduli of the elements of the $\tilde{O}$ matrix given by the statevector simulation for all the regimes studied on the QPU. The upper row corresponds to the large cluster, while the lower row corresponds to the small cluster; the number of qubits is indicated above each matrix, as well as the $J/h$ ratio. Each column corresponds to a distinct instance of the TFIM, with the first three columns covering the pure TFIM, the fourth covering the case with a longitudinal field, and the last covering the TFIM on a ladder geometry.}
    \label{fig:xelemssv}
\end{figure*}

 To give a sense of the difficulty in inverting the matrices of all the models that we have studied so far, we show in Fig.~\ref{fig:xelemssv} the matrix elements computed using the CX-test, that is the $\tilde{O}$ matrix elements. As we get closer to criticality, the matrix gets denser and further away from an easy inversion, since it grows out from the diagonal. To get an intuition about this difficulty, note that for a banded, invertible matrix, the inverse matrix elements scale with the bandwidth of the matrix~\cite{demko_1984}. In other words, the further away from a diagonal matrix, the more complex the inverse.

\section{Adiabatic sweep}\label{app:goodsweep}

Now, we discuss in more detail NLCE+ASP results for 1QP dispersion in the thermodynamic limit, results that were first shown in Fig.~\ref{fig:TFIM_30j_0hl_results_sweep} for TFIM chain at $J/h=0.3$ and $J/h=0.8$ with $N_{\mathrm{max}}=5$. Additionally, we show noisy simulations that suggest quantum hardware improvements are needed before ASP becomes an alternative to VQEs. 

To analyze the feasibility of adiabatic transformations, we show noisy simulations, with decreased noise levels, in order to get insight on the error rate required for the quantum device to be able to sustain ASP.

In doing so, we also varied the number of steps in the adiabatic sweep to uncover if a trade-off between accuracy of the sweep and depth of the circuit was possible. At the reference noise level, it is not possible, in the sense that both amounts are overwhelmed by noise. By time steps, we refer to the discretized steps of the sweep, and not the Trotter steps themselves, which are applied to each time step. Because the sweep unitary is time-dependent, one must both discretize the changing time variable and trotterize the instantaneous Hamiltonians. 

\begin{figure}[h!]
    \centering
    \includegraphics[width=.45\linewidth]{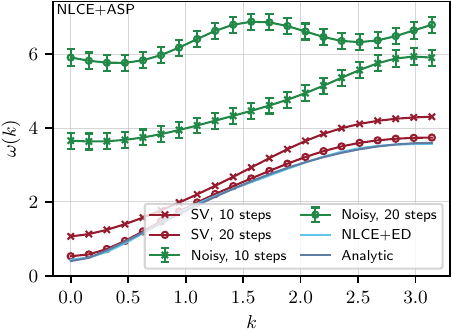}    \includegraphics[width=.45\linewidth]{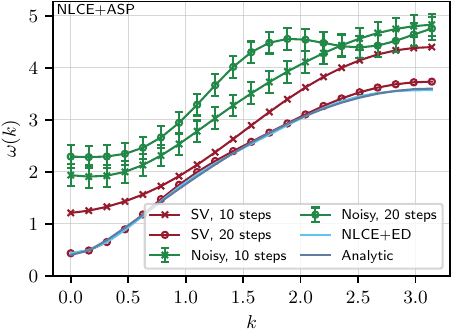}
    \includegraphics[width=.45\linewidth]{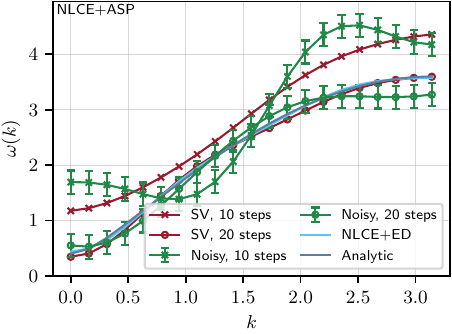}
    \caption{Same setting as in Fig.~\ref{fig:TFIM_80j_0hl_results_sweep} with $J/h=0.8$ for the one-dimensional TFIM; here the number of steps indicates how many time steps were used in the sweep.  \textbf{Top left:} Current error levels, \textbf{Top right:} with depolarizing noise divided by 10, \textbf{Bottom:} with depolarizing noise divided by 100. The error bars denote one standard deviation of statistical uncertainty.}
    \label{fig:aqtsweep_better}
\end{figure}

The dispersion curves in Fig.~\ref{fig:aqtsweep_better} very clearly indicate that an improvement in the error rate of one to two orders of magnitude is desirable for a satisfactory run of the adiabatic sweep, in the context of the NLCE+ASP framework, if we assume that the reference error rates match the hardware error rates, and given no other error mitigation technique. We note that, in different situations which might not involve non-linear processing of the quantum data, higher error rates might be tolerable.

In the bottom panel, with noise decreased a hundredfold, the noisy dispersion almost overlays the noiseless one, and the number of steps in the sweep becomes the limiting factor. Then, the main source of error is the accuracy of the algorithm, and not that of the device. We note that  the 20-step adiabatic sweep always brings us close to the analytic solution, whereas the 10-step sweep does not.

\begin{figure}[h]
    \centering
    \includegraphics[width=.95\linewidth]{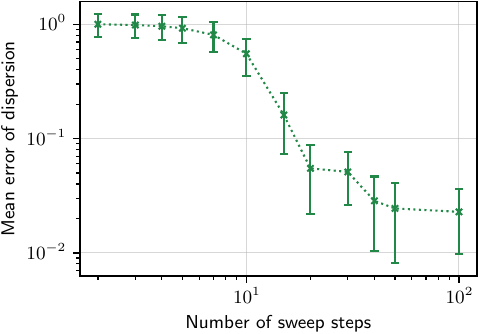}    
    \caption{Statevector simulation of NLCE+ASP, same setting as in Fig.~\ref{fig:TFIM_80j_0hl_results_sweep} with $J/h=0.8$ for the one-dimensional TFIM.}
    \label{fig:sweep_converge}
\end{figure}

Finally, we also show in Fig.~\ref{fig:sweep_converge} the mean error on the dispersion as the number of sweep steps, and therefore sweep time, increases, using a noiseless statevector simulation. As expected, very few steps do not suffice. The section between 10 and 20 steps clearly appears to be where the trade-off between accuracy and complexity is maximal. Note that beyond 20 steps, improvements in accuracy diminish. 

\section{Monte Carlo sampling}\label{app:mcconv}

\begin{figure}[h]
    \centering
    \includegraphics[width=.95\linewidth]{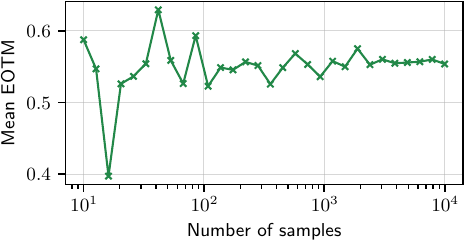}
    \caption{Convergence of the mean error-on-the-mean (EOTM) of the dispersion with respect to the number of Monte Carlo samples used. The dispersions are generated using the QPU uncertainty data obtained in Sec.~\ref{sec:results}, for a TFIM chain with $J/h=1$, using NLCE+VQE with $N_{\text{max}}=5$.}
    \label{fig:mcconv}
\end{figure}

In order to propagate the uncertainty of the expectation values through the NLCE+QA framework, we must perform Monte Carlo sampling with an appropriate number of samples. We choose to use $M_{\rm mc} = 10^4$ samples based on the plot in Fig.~\ref{fig:mcconv}, which is generated in the context of a NLCE+VQE framework for a TFIM model at $J/h=1$ and $N_{\text{max}}=5$. The mean EOTM is computed by generating many dispersions, computing the standard deviation at each point $k$ and then taking the average of these standard deviations, i.e.~$1/P \sum_l^P \sqrt{\mathbb{E}[\omega^2(k_l)] - \mathbb{E}[\omega(k_l)]^2}$. The number $P$ of $k_l$ points is chosen arbitrarily after the quantum data is collected. 

\section{The Hadamard test}\label{app:hadamtest}

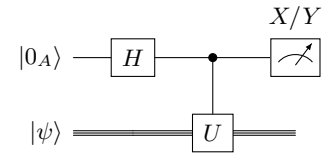
\begin{figure}[h]
    \centering
        \begin{quantikz}[wire types={q,b}, thin lines=true]
        \lstick{$\ket{0_A}$} &\gate{H}&\ctrl{1}&  \meter{X/Y} \\
        \lstick{$\ket{\psi}$} &        &\gate{U}         & \rstick{} 
        \end{quantikz}
    \caption{The Hadamard test. The first register, denoted by the $A$ index, has one qubit, whereas the second register with initial state $\ket{\psi}$ has multiple.}
    \label{fig:Hadamardtest}
\end{figure}

In this part, we write out a brief reminder of how to compute overlaps between quantum states with the well-known Hadamard test circuit.
For some unitary $U$ and some initial state $\ket{\psi}$, one may compute $\mathrm{Re}{\bra{\psi}U\ket{\psi}}$ and $\mathrm{Im}{\bra{\psi}U\ket{\psi}}$ with the Hadamard test, whose circuit is depicted in Fig.~\ref{fig:Hadamardtest}. 

In the Hadamard test, the ancilla in superposition is measured in the $X$-basis to obtain the real part, and its expectation value is
\begin{align}
    &\tfrac{1}{2}\bigl(\bra{\psi}U^\dagger\bra{1_A}
    + \bra{\psi}\bra{0_A}\bigr)
    (X \otimes I)
    \bigl(\ket{0_A}\ket{\psi}
    + \ket{1_A}U\ket{\psi}\bigr)
    \notag \\
    &= \tfrac{1}{2}\bigl[
    \braket{\psi|U^\dagger|\psi}
    + \braket{\psi|U|\psi}\bigr]
    \notag \\
    &= \mathrm{Re}\,\braket{\psi|U|\psi}.
\end{align}
 Measuring in the $Y$-basis gives $\mathrm{Im}\,\langle\psi|U|\psi\rangle$. In order to compute the off-diagonal elements of $U$, one may simply perform the Hadamard test on $\bra{0} X_i U X_j \ket{0}$. However, on noisy quantum hardware, implementing the controlled unitary $U$ is a challenge in terms of depth, especially if the connectivity is limited. 
 
 In this work, all circuits can be represented with time-evolution unitaries, implemented with Trotterized Pauli sum exponentials; the difference between the VQEs and sweeps being that the evolution times are parametrized in the former. On platforms with all-to-all connectivity, this results, using the standard decomposition of Pauli string exponentials to circuit~\cite{nielsen_Quantum_2010}, to all single-qubit $Z$-rotation gates becoming two-qubit controlled $Z$-rotations. On platforms with restricted connectivity, implementing the control may be more difficult and involve SWAP gates that increase the depth of the circuit further. 
 
\onecolumngrid

\end{document}